\documentclass[]{aa}
\def\apjs{ApJS}
\def\aap{A\&A}

\def\ueber#1#2{{\setbox0=\hbox{$#1$}%
  \setbox1=\hbox to\wd0{\hss$ #2$\hss}%
  \offinterlineskip
  \vbox{\box1\box0}}{}}

\def\lapprox{\,\lower 1mm \hbox{\ueber{\sim}{<}}\,}
\def\gapprox{\,\lower 1mm \hbox{\ueber{\sim}{>}}\,}

\begin{document}

\title{The  late   stages  of  the   evolution  of  intermediate--mass
       primordial stars: the effects of overshooting}

\author{Pilar Gil--Pons\inst{1},
        Jordi Guti\'errez\inst{1} \and 
        Enrique Garc\'{\i}a--Berro\inst{1,2}}

\offprints{E. Garc\'\i a--Berro}

\institute{Departament de F\'\i sica Aplicada, Escola  Polit\'ecnica 
           Superior de Castelldefels, Universitat   Polit\`ecnica de 
           Catalunya,   Avda.  del  Canal  Ol\'\i  mpic  s/n,  08860 
           Castelldefels,    Spain     (e-mail:     jordi,    pilar, 
           garcia@fa.upc.es)\
           \and
           Institute for Space Studies of Catalonia, c/Gran Capit\`a
           2--4, Edif.  Nexus 104, 08034 Barcelona, Spain}

\date{\today}

\abstract{}  
         {We compute and analyze  the evolution of primordial stars of
         masses  at  the ZAMS  between  $5  \,  M_{\sun}$ and  $10  \,
         M_{\sun}$, with and without overshooting.  Our main goals are
         to  determine   the  nature   of  the  remnants   of  massive
         intermediate--mass   primordial  stars   and  to   check  the
         influence of overshooting in their evolution.}
         {Our calculations  cover  stellar  evolution  from  the  main 
         sequence  phase until the  formation  of the degenerate cores  
         and the thermally pulsing phase.}
         {We  have obtained  the  values for  the  limiting masses  of
         Population III progenitor stars leading to carbon--oxygen and
         oxygen--neon compact cores.   Moreover, we have also obtained
         the limiting  mass for which isolated  primordial stars would
         lead to  core--collapse supernovae after the end  of the main
         central  burning phases.   Considering a  moderate  amount of
         overshooting  the   mass  thresholds  at  the  ZAMS  for  the 
         formation of carbon--oxygen and oxygen--neon degenerate cores  
         shifts  to  smaller  values  by  about  $2\, M_{\sun}$.  As a 
         by--product of our  calculations, we  have also  obtained the  
         structure  and  composition profiles of the resulting compact 
         remnants.}
         {Opposite to what happens with solar metallicity objects, the
         final  fate  of  primordial  stars   is  not  straightforward
         determined from the  mass of the compact cores  at the end of
         carbon burning. Instead, the small mass--loss rates typically
         associated to  stellar winds  of low metallicity  stars might
         allow the growth of the  resulting degenerate cores up to the
         Chandrasekhar  mass, on  time  scales one  or  two orders  of
         magnitude  shorter  than  the  time  required  to  loose  the
         envelope.  This would lead to the formation of supernovae for
         initial masses  as small as  $\sim 5 \, M_{\sun}$.}

\titlerunning{The late  stages of the  evolution of intermediate-mass
              primordial stars.}  
\authorrunning{Gil--Pons et al.}

\maketitle


\section{Introduction}

Primordial stars ---  that is, stars formed in  the early Universe ---
are direct  inheritors of the  matter synthesized during the  Big Bang
(Alpher  \& Herman, 1950;  Olive, 2000).   The initial  composition of
these objects --- which is  characterized by the absence of metals ---
determines their evolution and, ultimately, the fundamental properties
of  the  resulting  stellar  remnants,  as  well  as  the  amount  and
composition of  the matter returned  to the interstellar  medium after
the main evolutionary phases.  Therefore, the evolution of these stars
is  important  for  a  correct  modeling  of  the  population  of  the
degenerate remnants of  Population III stars and, hence,  for a better
understanding of the distribution of baryonic dark matter, whereas the
significance  of  the  ejected  material  rests on  its  influence  on
Galactic chemical evolution.

Even though  the detection of primordial composition  objects --- that
is stars with  [Fe/H]~$\lapprox -8.3$ --- has not  been possible up to
now, during  the last few years several  encouraging observations have
been made.  For instance, Bessel \& Norris (1984) detected a red giant
with [Fe/H]~$\approx -4.6$ and  Christlieb et al.  (2002) measured the
metallicity  of HE~0107--5240,  a low  mass star  with [Fe/H]~$\approx
-5.3$.   Even  more recently,  Frebel  et  al.   (2005) have  observed
HE~1327--2326, an  unevolved low-mass object, which was  found to have
[Fe/H]~$\approx -5.4$.   Consequently, both the  intrinsic theoretical
interest and  the growing observational evidence for  the existence of
primordial stars  make the study of  these objects one  of the hottest
topics in stellar astrophysics nowadays.

The exact shape  of the initial mass function  of Population III stars
is far from being known. The standard view advocates for a strong bias
toward very  massive stars, say  between 100 and  $600\,M_{\sun}$ ---
see,  for  example,  Bromm,  Coppi \&  Larson  (2002).   Nevertheless,
arguments  have been  put forth  recently that  point to  the possible
existence  of a  population  of very  metal  poor stars  of low--  and
intermediate--mass.  For instance, from the theoretical side, Nakamura
\& Umemura (1999, 2001) have  simulated the growth of instabilities in
protostellar primordial  clouds in 2 and 3  dimensions.  Their results
indicate that the initial mass function could be bi--modal with a peak
at $\sim 100\,  M_{\sun}$ and a secondary peak at  a few solar masses.
Additionally, Machida et al.  (2005)  have shown that the explosion of
primordial supernovae can drive  the Jeans masses of primordial clouds
from a few $10^6\,M_{\sun}$  to values well below $1\,M_{\sun}$, hence
allowing  for  the formation  of  intermediate--mass  stars, but  with
non--negligible metallicities (of the order of [Fe/H]~$\approx -3.0$).
Furthermore,  Johnson  \&  Bromm  (2006)  have  recently  raised  more
theoretical  considerations   that  suggest  that   the  formation  of
primordial low-- and intermediate--mass  stars is viable. On the other
hand, and from the observational side, Schneider et al. (2006), on the
basis of three tentative NICMOS  UDFs sources at $z\approx 10$ and the
derived  WMAP  electron scattering  optical  depth,  predict a  rather
standard form of the IMF, with stars within the $1\,M_{\sun} \la M \la
100 \,M_{\sun}$ range.

The evolution of  zero metallicity stars has been  analyzed in several
recent publications. Just to mention  some of them, and without trying
to be complete,  it is important to realize that  Heger et al.  (2001)
and Heger \&  Woosley (2000) have focused on  massive objects, whereas
Limongi et al. (2000), Marigo et al. (2001), Chieffi et al. (2001) and
Siess et al. (2002) have thoroughly studied the evolution of low-- and
intermediate--mass   stars.   However,   the   evolution  of   massive
intermediate--mass primordial  stars has been  largely ignored, except
for  the case  of the  evolutionary sequence  of a  $9\,M_{\sun}$ star
presented in Gil--Pons et al. (2005). Probably, one of the reasons for
this  is  the  heavy  computational  load involved  in  computing  the
evolution during the carbon burning phase for this mass range (Ritossa
et al. 1995; Garc\'\i a--Berro et al. 1997; Iben et al.  1997; Ritossa
et al. 1999).  This is in  sharp contrast with the situation for stars
of  Population I  and II,  that  have been  thoroughly analyzed.   For
instance, the mass thresholds for objects that lead to different types
of  remnants have been  determined for  Population I  and II  stars by
Heger  et al.   (2003), as  well as  by Eldridge  \& Tout  (2004), who
computed a rich grid of stellar models ranging both in mass --- from 5
to $200 \, M_{\sun}$ --- and  metallicity --- from $Z = 10^{-5}$ up to
$Z=0.05$.   The latter  authors also  payed special  attention  to the
adopted   mass--loss  rates,  including   their  variation   with  the
metallicity, and to the effects of overshooting.  Finally, it is worth
mentioning as well  that the influence of Population  III stars on the
formation  and ejection of  isotopes to  the intergalactic  medium has
been extensively  studied by  Goriely \& Siess  (2002) and by  Abia et
al. (2001).

In  this paper,  we  describe the  final  stages of  the evolution  of
massive intermediate mass primordial  stars.  Reliable results on this
matter are essential inputs for Galactic chemical evolution models, as
well  as  for   the  supernova  theory.   Particularly,  thermonuclear
supernovae (SNeIa),  in which  a carbon--oxygen white  dwarf overcomes
the Chandrasekhar limit, must rely on accurate initial--to--final mass
relations to reproduce the observed supernova rate, and its dependence
on  metallicity could  represent a  source of  diversity  on classical
SNeIa at different redshifts.   Specifically, we outline the evolution
of the  progenitor stars and  determine the mass thresholds  to obtain
carbon--oxygen  (CO),  oxygen--neon (ONe)  degenerate  cores and  core
collapse supernovae at the end of the main central burning stages. The
stars hosting CO  degenerate cores could give raise to  CO novae or to
SNeIa  if  belonging to  a  close  binary  system; those  hosting  ONe
degenerate  cores could  also  form  ONe novae  or  supernovae by  the
accretion--induced  collapse  mechanism   (Canal  \&  Schatzman  1976;
Guti\'errez  et al.   1996; Guti\'errez,  Canal \&  Garc\'\i a--Berro,
2005), and core--collapse supernovae would directly form from isolated
stars when nuclear burning proceeds all the way to the formation of an
iron peak element dominated core.

An  important part  of our  results  consists in  justifying that  the
formation of supernovae does not require ZAMS masses as high as in the
case of  solar metallicity objects but, instead,  $6\, M_{\sun}$ stars
(and  probably  even  less  massive  ones) are  enough  to  produce  a
supernova explosion. This possibility  had already been suggested, for
instance,  by Arnett  (1969), by  Iben \&  Renzini (1983),  by Willson
(2000) and, more  recently, by Zijlstra  (2004).  The reason  for such
peculiar behavior has to do  with the characteristics of stellar winds
in  primordial stars.   Low metallicity  stars are  supposed  to drive
weaker stellar winds (Bowen \&  Willson 1991; Willson 2000) than their
solar  metallicity counterparts.  However,  some caution  is mandatory
here,  as  the  cited  works  are  theoretical  with  phenomenological
character, and observations  do not fully support ---  nor discard ---
their findings yet.  Therefore, the  degenerate cores are able to grow
significantly due to shell burning  before the envelope is lost. As we
will show  below, the  cores resulting from  the evolution  of massive
intermediate--mass primordial  stars will  be able to  grow up  to the
Chandrasekhar mass and, therefore, undergo a supernova explosion.

Of  course,  all  these  results  depend sensitively  on  the  adopted
prescription  for convection  and,  in particular,  on  the degree  of
overshooting.   Overshooting represents  one of  the main  unknowns in
stellar evolution, as observations indicate more mixing than predicted
by standard, local  and non--local convective prescriptions.  Rotation
and  purely three--dimensional  hydrodynamic  phenomena could  explain
this extra mixing, but this is still the subject of an active, ongoing
debate.  Hence,  we have performed calculations both  with and without
overshooting in order to clarify its effects on the initial--to--final
mass relation.   Previous works on overshooting ---  see, for example,
Young  et  al.   (2003)  and  Demarque  et  al.   (2004)  for  recent,
enlightening  studies  ---  indicate  that  overshooting  produces  an
increase in  the mass of the  nuclear exhausted core, a  change of its
internal  composition,  and a  modified  chemical  composition of  the
surface layers.   All these effects  are relevant for the  purposes of
our  work. In particular  it is  of the  maximum importance  the extra
mixing  in  the  surface  layers,  as  population  III  stars  do  not
experience a significant mass--loss due  to the small opacity of their
outer layers.  The extra mixing due to overshooting could increase the
metallicity  of  the  outer  layers and,  consequently,  increase  the
mass--loss rates, thus changing the final mass of the remnants.

The paper is organized as follows.  In Sect. 2 we describe the overall
evolution prior  to carbon burning. Section  3 is devoted  to study in
depth the  different carbon burning episodes.   In \S 4  we comment on
the thermally  pulsing phase for this  mass range, whereas in  \S 5 we
describe  the properties  of the  cores of  these stars.   Finally, in
section 6 we elaborate on  the possible evolutionary outcomes that can
be envisaged for  these stars, discussing in depth  the effects of the
adopted mass--loss rates,  whilst in section 7 we  summarize our major
findings, we draw our conclusions and we discuss their significance.

\begin{table*}
\centering
\caption{Main nuclear burning times and core masses.}
\begin{tabular}{ccccccccc}
\hline
\hline
  \multicolumn{1}{c}{$M_{\rm ZAMS}/M_{\sun}$}
& \multicolumn{2}{c}{$t_{\rm CHB}\,(\times 10^{14}$~s)}
& \multicolumn{2}{c}{$M_{\rm He}/M_{\sun}$}
& \multicolumn{2}{c}{$t_{\rm CHeB}\,(\times 10^{14}$~s)}
& \multicolumn{2}{c}{$M_{\rm CO}/M_{\sun}$}
\\
\cline{2-9}
 & no OV & OV  & no OV & OV  & no OV & OV  & no OV & OV \\
\hline
 5.0 & 19.8838 & 21.2417 & 1.17 & 1.32 & 23.6791 & 24.2174 & 0.64  & 0.80 \\
 6.0 & 13.8396 & 15.3209 & 1.18 & 1.60 & 16.2184 & 17.0675 & 0.65  & 0.90 \\
 7.0 & 10.5091 & 11.8733 & 1.20 & 2.23 & 12.1470 & 12.8494 & 0.83  & 1.15 \\
 8.0 &  8.3796 &  9.4119 & 1.48 & 2.30 &  9.7551 & 10.3818 & 1.09  & 1.38 \\
 9.0 &  6.8994 &  7.7360 & 1.85 & 2.35 &  8.0519 &  8.5778 & 1.18  & 1.50 \\
10.0 &  6.0116 &  6.7436 & 2.09 & 2.38 &  6.8359 &  7.3791 & 1.30  & 1.76 \\
\hline
\hline
\end{tabular}
\end{table*}


\section{Overall evolution prior to carbon burning}

\begin{figure}[t]
\vspace{7.9cm}
\hspace{-2.7cm}
\includegraphics{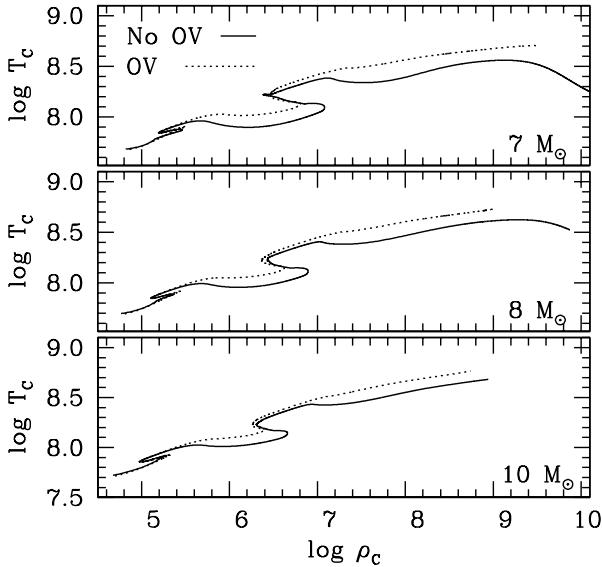}
\caption{Evolution in the $\log \rho_{\rm c} -\log T_{\rm c}$ plane of
        our model stars  up to the point where  carbon is ignited. The
        solid lines  correspond to the  case in which  no overshooting
        was taken into account, whereas the dotted lines correspond to
        the calculations in which overshooting was incorporated.}
\end{figure}

\begin{figure}[t]
\vspace{7.9cm}
\hspace{-2.7cm}
\includegraphics{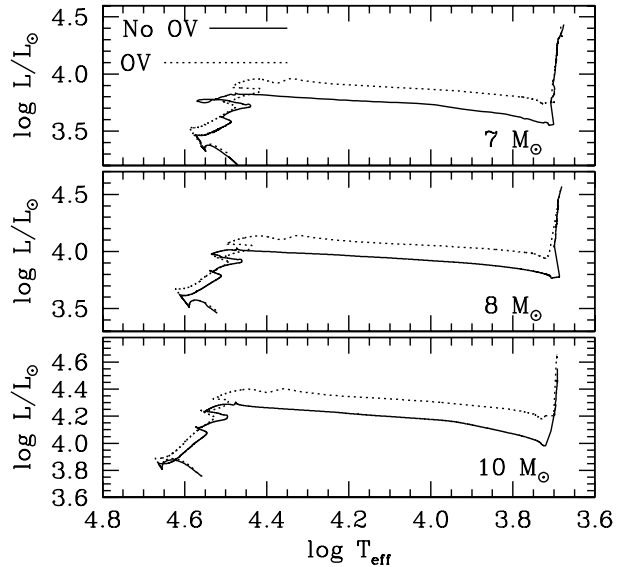}
\caption{Evolutionary  tracks in  the Hertzsprung--Russell  diagram of
         our model  stars. The solid  lines correspond to the  case in
         which overshooting was  disregarded, whereas the dotted lines
         show the evolution for the case in which overshooting was due
         taken into account.}
\end{figure}

We have followed the evolution of zero--metallicity models with masses
at the  ZAMS between 5  and $10 \,  M_{\sun}$, and initial  values for
hydrogen and  helium abundances of  {\rm $X=0.77$} and  {\rm $Y=0.23$}
respectively. Our  calculations have  been performed both  taking into
account and  neglecting overshooting.  In  doing so, we have  used the
evolutionary code  previously described  in Gil--Pons et  al.  (2005),
which  is,  itself, an  update  of the  code  described  in detail  in
Ritossa,  Garc\'{\i}a--Berro  \&  Iben  (1995).  We  remark  that  the
opacities are obtained  by interpolating in the tables  of Iglesias \&
Rogers  (1993).   The  only  novelty  with  respect  to  the  previous
implementations of our code is that for the calculations reported here
we have incorporated  overshooting. This will allow us  to isolate the
differences introduced  by the  treatment of convective  boundaries in
the  evolutionary  calculations  presented  here.   Specifically,  the
prescription used for  the calculation of the convective  edges is the
one  described in  Eldridge  \&  Tout (2004),  in  which the  standard
Schwarzschild  criterion is  modified to  let convection  be  the main
energy  transport   mechanism  in  those  zones  of   the  star  where
$\nabla_{\rm rad} > \nabla_{\rm ad} - \delta$, with

\begin{equation}
\delta = \frac{\delta_{\rm OV}}{2.5+20 \zeta + 16 \zeta^2}
\end{equation}

\noindent where $\zeta=P_{\rm  rad}/P_{\rm gas}$ and $\delta_{\rm OV}$
is set to be equal to 0.12 (Schr\"oder, Pols \& Eggleton 1997).

The evolution  prior to carbon  burning follows very closely  the main
trends previously described in Gil--Pons  et al. (2005). That is, core
hydrogen  burning  (CHB) begins  through  the  pp--chains until  small
amounts of  carbon ($X($C$)  = 10^{-10}$ by  mass) form and  allow the
onset of the CNO cycle.  The  rest of the CHB phase proceeds therefore
at the higher central temperatures that characterize this cycle.  Once
CHB has ended,  core contraction proceeds and the  conditions for core
helium burning are  reached in the central regions  of the star. Those
for  hydrogen  burning are  attained  at  the  layers just  above  the
hydrogen  exhausted  core.   Opposite   to  what  happens  with  solar
metallicity stars, all  these phases take place at  the blue region of
the Hertzsprung--Russell diagram.  Furthermore, hydrogen shell burning
in  a metal--free  environment  is  not able  to  produce the  overall
expansion  of the stellar  envelope that  characterizes the  red giant
branch.

\begin{figure}[t]
\vspace{7.9cm}
\hspace{-2.7cm}
\includegraphics{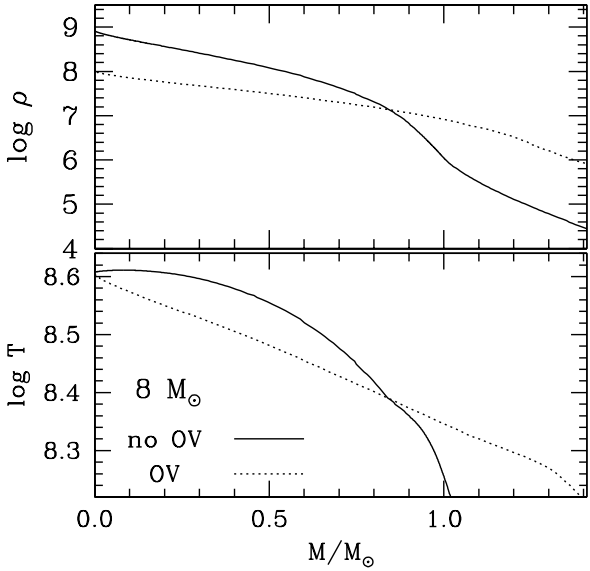}
\caption{Upper panel: density profile of the degenerate carbon--oxygen
         cores  for  the  $8\,  M_{\sun}$  model star  just  before  C
         ignition  for the case  computed without  overshooting (solid
         line) and  with  overshooting   (dotted  line).  Lower  panel:
         temperature  profiles of  the  carbon--oxygen cores  computed
         without  overshooting  (solid  line)  and  with  overshooting
         (dotted line) for the same model star.}
\end{figure}
         
As mentioned in Gil--Pons et al.  (2005), the main differences between
the evolutionary  tracks for $Z=0$  and those of $Z=Z_{\sun}$  are the
longer time spent  during the main sequence phase,  the higher surface
luminosities reached during CHB  and the larger helium--exhausted core
of the  $Z=0$ models.   Table 1  shows the duration  oh the  CHB phase
($t_{\rm CHB}$), the mass of the resulting helium core ($M_{\rm He}$),
the time at  which carbon burning begins ($t_{\rm  CB}$), and the mass
of  the  corresponding  carbon--oxygen  core.   As can  be  seen,  the
calculations  with overshooting  yield time  scales for  core hydrogen
burning  that are  about  a 12\%  longer  when compared  to the  cases
computed without  taking into account overshooting.   The helium cores
after the CHB phase has been  completed are also about 50\% larger and
this  is the reason  why the  core helium  burning (CHeB)  phase lasts
about 6\% longer in the  cases in which overshooting was incorporated.
Finally, the  masses of  the CO cores  $(M_{\rm CO}$) prior  to carbon
burning are  considerably larger  as well, that  is, about 25  \% more
massive for the cases computed with overshooting.

Figure 1  shows the  evolution of the  central temperature  versus the
central density for both sets of models. The solid lines correspond to
the cases in  which overshooting was disregarded and  the dotted lines
to the  cases in which overshooting  was taken into  account.  For the
sake of conciseness we only  show the evolutionary sequences of the 7,
8  and  $10\, M_{\sun}$  model  stars.  As  can  be  seen, only  small
differences  appear  between  both  sets  of  models  during  the  CHB
phase. However,  the core contraction  phases that follow the  CHB and
the  CHeB phases  take place  at higher  temperatures and  reach lower
values  for the  maximum central  density in  the cases  computed with
overshooting.

\begin{figure}[t]
\vspace{7.9cm}
\hspace{-2.7cm}    
\includegraphics{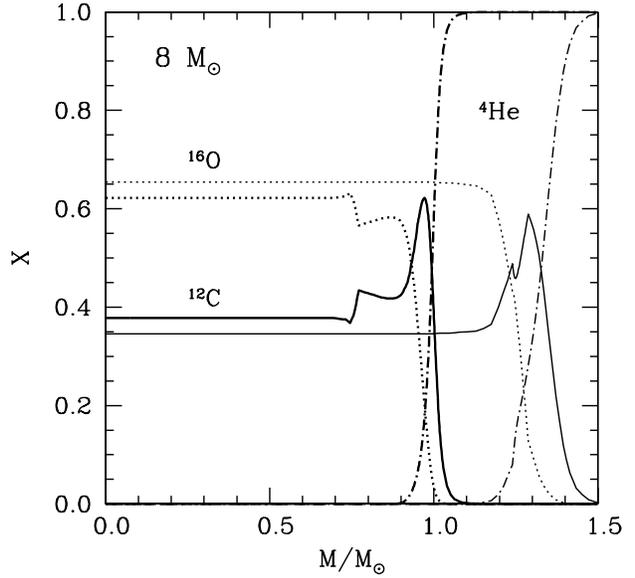}
\caption{Composition profiles  of the degenerate CO cores  of the $8\,
         M_{\sun}$ model star  computed with overshooting (thin lines)
         and  disregarding  overshooting  (thick  lines)  just  before
         carbon burning sets in.}
\end{figure}

\begin{figure*}[t]
\vspace*{10.8cm}
\hspace*{-2.7cm}    
\includegraphics{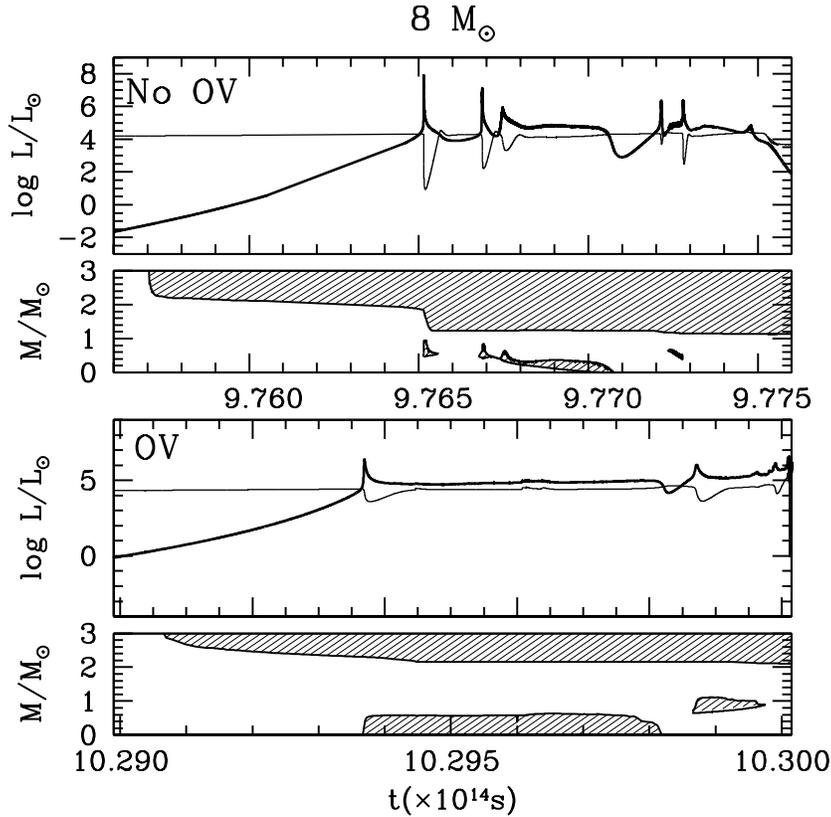}
\caption{Temporal evolution  of the main  structural parameters during
        the bulk  of the carbon  burning phase for the  $8 \,M_{\sun}$
        model computed without overshooting  (upper panel) and with it
        (bottom panel).   The shaded areas represent  the evolution of
        the  convective  regions, the  thick  solid  line depicts  the
        evolution of the carbon  burning luminosity and the thin solid
        line shows the helium burning luminosity. }
\end{figure*}
                                             
The evolution  in the Hertzsprung--Russell  diagram of the  same model
stars is shown in figure 2. In  the upper panel the track of the $7 \,
M_{\sun}$  star  is  shown, and  in  the  middle  and lower  panel  we
represent the  evolution of the $8  \, M_{\sun}$ and  $10 \, M_{\sun}$
models, respectively.  The cases  with no overshooting are represented
as solid lines  and the cases with overshooting  are represented using
dotted lines.   As can  be seen, CHB  takes place at  higher effective
temperatures  for  the  models  computed with  overshooting,  and  the
Hertzsprung gap occurs at larger luminosities.  In summary, the models
computed with  overshooting behave as  if they were more  massive than
their counterparts computed without overshooting.

This can be  better seen by comparing the masses of  the CO cores just
before  the carbon  burning phase  (Table  1). Also,  the density  and
temperature profiles  just before carbon ignition ---  which are shown
in Fig.  3  for the $8 \, M_{\sun}$ model star  --- help in displaying
this  distinctive  behavior.  Particularly,  the  model computed  with
overshooting  presents  a  smaller  central density  and  a  shallower
density  profile --- see  the top  panel of  Fig. 3  --- leading  to a
larger region of high densities  and, therefore, to a larger core than
the  model computed  without taking  into account  overshooting. Note,
however,  that  the central  temperatures  (see  the  bottom panel  of
Fig. 3)  of both models are  remarkably similar, although  the mass of
the core is rather different.  Figure 4 shows the chemical profiles of
the same model star. The  thick lines correspond to the model computed
with  no overshooting  and  the  thin lines  correspond  to the  model
computed  when  due account  is  taken  of  overshooting.  The  former
appears to be  $0.3 \, M_{\sun}$ larger than the  latter, but the mass
abundances of carbon and oxygen in the central regions of the star are
practically  the  same  in  both  cases.   All  these  quantities  are
particularly  important because,  as it  will  be shown  in section  4
below, the  development of  extensive carbon burning  and, ultimately,
the formation of an oxygen--neon white dwarf crucially depend on them.
The mass  threshold at the ZAMS  for the formation  of supernovae from
single  $Z  = 0$  stars  also  depends  significantly on  the  adopted
prescription for determining the edge of the convective cores.


\section{The carbon burning phase}

Figure  5  represents  the  temporal  evolution of  the  base  of  the
convective envelope  (BCE) as well  as the luminosities  associated to
carbon burning (thick solid line) and helium burning (thin solid line)
for the $8 \, M_{\sun}$ model stars computed without overshooting (top
panel) and with overshooting (bottom panel), which is a representative
case.  The  convective regions associated  to carbon burning  are also
shown for both  models. We remark at this point  that neither the $6\,
M_{\sun}$  nor  the  $7\,   M_{\sun}$  model  stars  computed  without
overshooting develop  carbon burning extensively ---  although for the
latter  case there  is  partial carbon  burning,  reaching the  carbon
luminosity  a maximum  value of  only $L/L_{\sun}\simeq  10$  --- and,
therefore, the remnant cores are composed mainly of carbon and oxygen.
In fact, our calculations show that the mass threshold at the ZAMS for
effective carbon burning in the partially degenerate core to obtain an
ONe  remnant is  $\approx 7.8  \, M_{\sun}$,  when no  overshooting is
considered.  This mass  threshold is  $\approx 6.0  \,  M_{\sun}$ when
overshooting is taken into account.

The  first important  feature  to be  noticed  in Fig.~5  is the  fast
advance of the base of  the convective envelope that, for both models,
occurs shortly after the end of  the core helium burning phase and the
onset  of helium  burning  in a  shell  --- at  $t\simeq 9.757  \times
10^{14}$~s for the model computed without overshooting and at $t\simeq
10.2907\times  10^{14}$~s for  the model  computed  with overshooting.
This  first  advance  inward,  the  so--called  ``second''  dredge--up
episode  ---  although,  in  fact,  is the  first  dredge--up  episode
(Gil--Pons et al. 2005) --- is  followed by a very slow penetration of
the BCE that finally allows for a moderate enrichment in metals of the
hydrogen--rich  envelope.  Although most  of the  material dredged--up
consists of helium  which does not contribute to  the metal content of
the  convective  envelope some  small  amounts  of  metals are  indeed
dredged--up.  The  metallicity of the envelope  depends sensitively on
the mass of the star. For instance, the metallicity of the envelope of
the  $5  \, M_{\sun}$  model  star  computed  without overshooting  is
$Z_{\rm env} \approx 10^{-9}$, whereas for the $9 \, M_{\sun}$ star we
obtain  $Z_{\rm  env}  \approx  10^{-4}$.  The  models  computed  with
overshooting show the same behavior: $Z_{\rm env} \approx 10^{-7}$ for
the $5 \, M_{\sun}$ star and  $Z_{\rm env} \approx 10^{-3}$ for the $8
\, M_{\sun}$.  As it will be  shown below, this enrichment plays a key
role in the final fate of massive intermediate--mass primordial stars.
The  reason  for this  behavior  can  be  understood by  studying  the
evolution  of  the  $8   \,  M_{\sun}$  model  star  computed  without
overshooting,  which shows  a particular  feature associated  with the
inner advance  of the convective  envelope.  For this model  the first
carbon flash  at $t\sim 9.766\times 10^{14}$~s occurs  near the center
of  the star,  but the  associated convective  zone extends  to layers
relatively close to the BCE (at $M_r\simeq 0.9 \,M_{\sun}$) and, thus,
allows for  a decrease  of the degeneracy  parameter of  the outermost
layers  of  the  core.   Consequently,  the  density  and  temperature
barriers decrease  and this  results in an  additional advance  of the
BCE.  This, in  turn, causes a enrichment in  metals of the convective
envelope.  For  the model computed  with overshooting this  feature is
absent, since carbon  ignition occurs at the center,  too far from the
edge of the carbon--oxygen partially degenerate core (which is located
at  $M_{\rm  BCE}\simeq  2.12  \,  M_{\sun}$)  to  have  any  relevant
consequence in the behavior of the convective envelope.

The  $8\,  M_{\sun}$  model  computed  without  overshooting  actually
undergoes 5 significant carbon flashes, the first of which is the most
powerful one ($\sim 10^8\,  L_{\sun}$). Once the overall degeneracy of
the core decreases  after this flash, the following  ones proceed with
more moderate luminosities, as it usually occurs for solar metallicity
models of equivalent  masses during this phase.  As  it also occurs in
solar  metallicity stars of  the same  mass, each  thermonuclear flash
generates an associated convective shell and an abrupt decrease of the
He  luminosity  (see Fig.~5).   This  is  a  consequence of  the  fast
injection  of  energy  in   the  uppermost  layers  of  the  partially
degenerate core,  which consequently  expand and cool  and, therefore,
effectively switch--off the He  burning shell.  For the model computed
taking  overshooting  into  account,  carbon ignition  occurs  at  the
center.  This  is a consequence  of the smaller central  densities and
lower temperatures achieved before carbon ignition (see Fig. 3), which
lead  respectively to a  smaller degeneracy  parameter and  to smaller
neutrino leaks. Consequently, the  temperature profile shows a central
maximum, in contrast  with the situation for the  $8\, M_{\sun}$ model
computed  without  overshooting,  for  which the  maximum  temperature
occurs relatively far from the  center. Additionally, we do not obtain
as many flashes  as for the model in which  overshooting was not taken
into account.  In fact, the  first carbon flash has a smaller strength
($L_{\rm C}\simeq  10^7\, L_{\sun}$).  This  also stems from  the fact
that the degeneracy  in the core is substantially  smaller.  After the
sudden  injection of  nuclear  energy,  the core  of  this model  star
expands and the overall degeneracy decreases, leading to a more gentle
carbon burning  phase, in which  carbon is burnt in  almost stationary
conditions.  Moreover, since the  outer edge of the central convective
carbon burning region  is located deep in the core,  the effect on the
He burning  shell is substantially  smaller and, consequently,  the He
burning luminosity decreases less than in the previous case.

\section{The thermally pulsing phase}

\begin{figure*}[t]
\vspace*{13cm}
\hspace*{-2.7cm}
\includegraphics{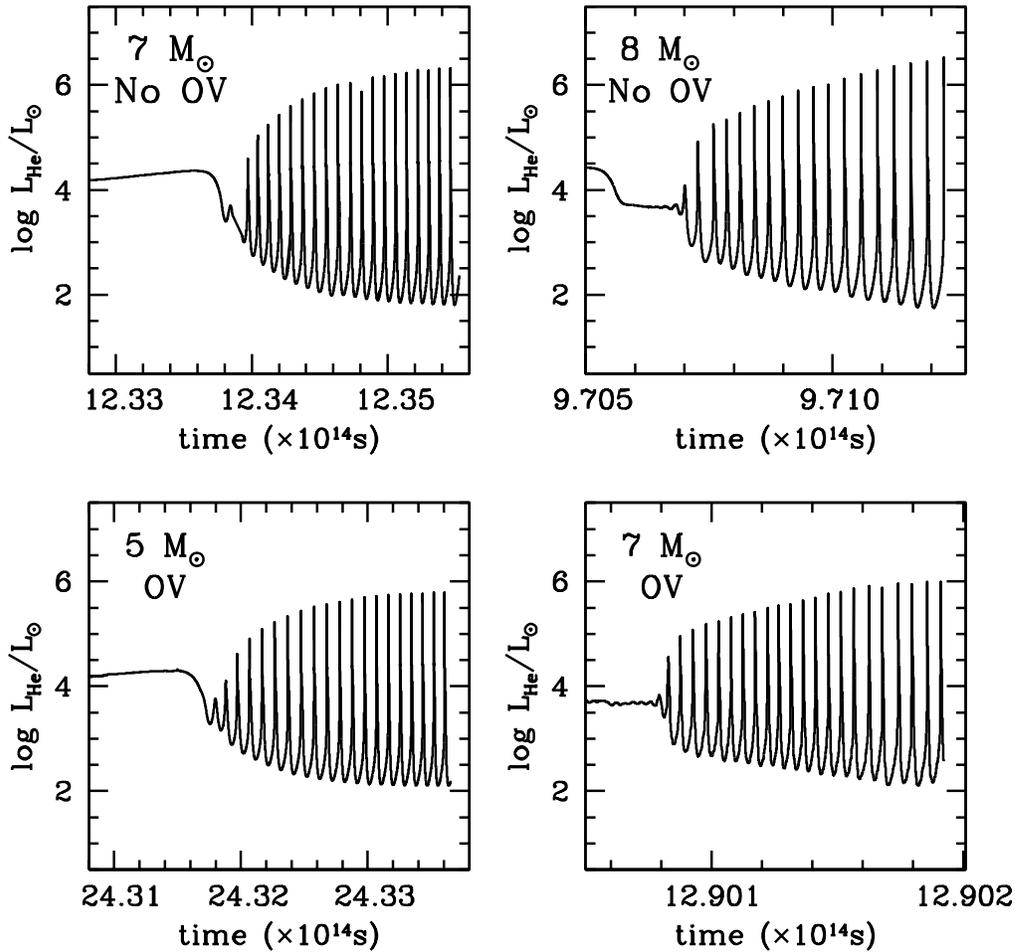}
\caption{Temporal   evolution  of  $L_{\rm   He}$  during   the  early
        TP--(S)AGB phase of the $7 \, M_{\sun}$ model computed without
        overshooting (upper  left panel), the $8  \, M_{\sun}$ without
        overshooting (upper  right panel),  the $5 \,  M_{\sun}$ model
        computed with  overshooting (lower left panel), and  the $7 \,
        M_{\sun}$ with overshooting (lower right panel).}
\label{pulses}
\end{figure*}

For the models which fail  to ignite carbon under partially degenerate
conditions a thermally pulsing  phase ensues soon after helium burning
ceases in  the central regions.  For these  models the helium--burning
shell  is very  close to  the  base of  the hydrogen--rich  convective
envelope.   For  instance,  for  the  $7\,  M_{\sun}$  model  sequence
computed without overshooting the  helium--burning shell is only $\sim
4.1\times  10^{-4} \,  M_{\sun}$  away from  the H--He  discontinuity.
Consequently, the  temperature at this discontinuity  increases to the
extent that hydrogen--burning is reactivated and the thermally pulsing
AGB phase  (TP--AGB) phase ensues  shortly afterward.  For  the models
which  do ignite  carbon the  thermally pulsing  Super--AGB (TP--SAGB)
phase ensues after the carbon  burning phase has been completed. This,
for instance,  is the case of  the $8\, M_{\sun}$  model star computed
without overshooting.   Again, in this case  the helium--burning shell
(located  at $M_r\simeq  1.049875 \,M_{\sun}$)  is very  close  to the
H--He   discontinuity  (located  at   a  mass   coordinate  $M_r\simeq
1.08169596\,M_{\sun}$)  and the hydrogen  burning shell  resurrects as
well.  Note, however, that Yoon et al. (2004) have recently shown that
other  parameters,  like  rotation,  define  the  stability  of  shell
burning.

\begin{figure*}[t]
\vspace*{12.5cm}
\hspace*{-1.7cm}
\includegraphics{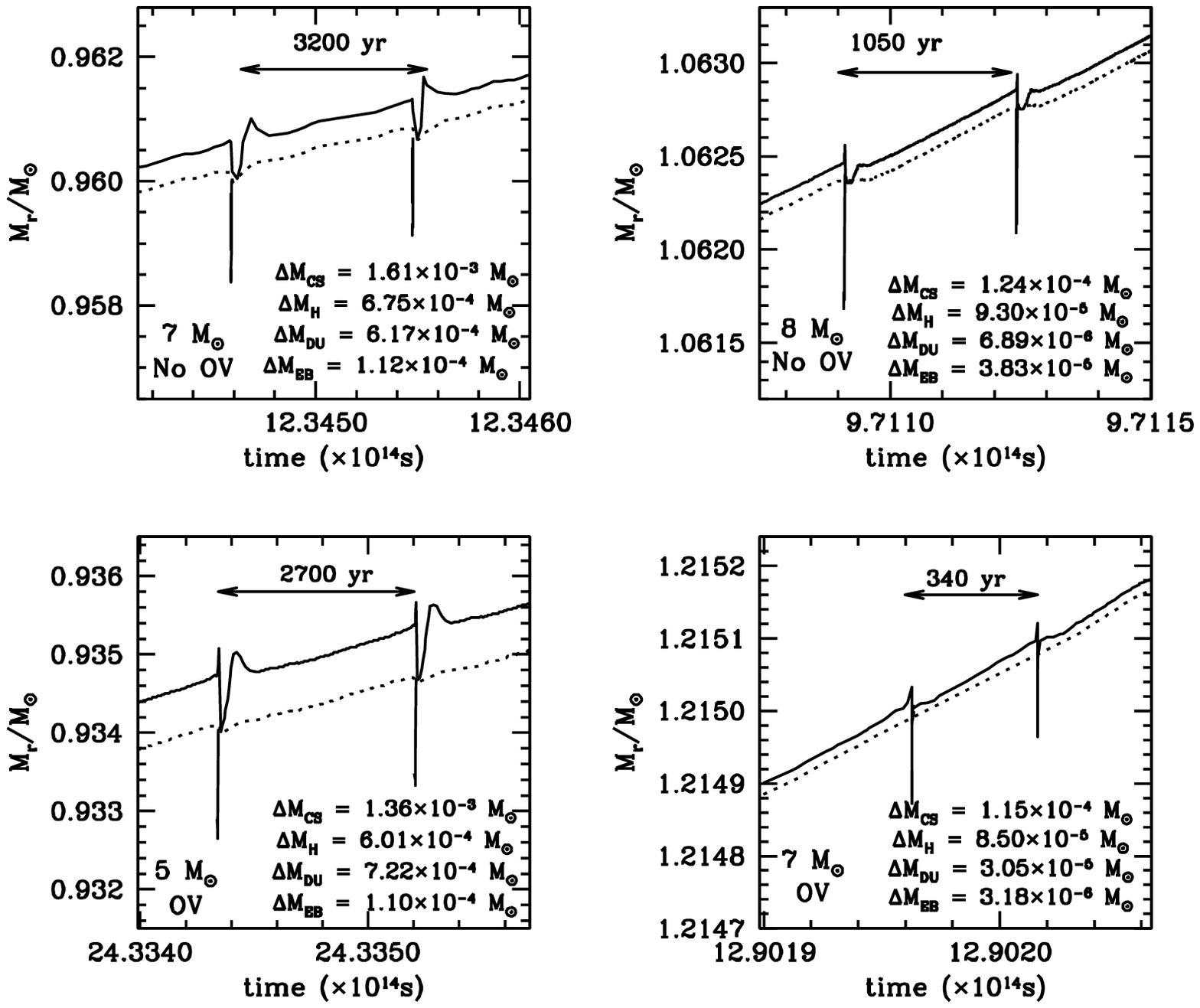}
\caption{Temporal evolution  of the edges of the  convective shell and
        of the base of the convective envelope during the ${\rm 8^{\rm
        th}}$ and  the ${\rm  9^{th}}$ pulses of  the $7  \, M_{\sun}$
        model star  computed without overshooting  (upper left panel),
        the ${\rm 14^{\rm th}}$ and  the ${\rm 15^{\rm th}}$ pulses of
        the $8 \, M_{\sun}$  star computed without overshooting (upper
        right panel),  the ${\rm 18^{\rm  th}}$ and the  ${\rm 19^{\rm
        th}}$  pulses  of the  $5  \,  M_{\sun}$  model computed  with
        overshooting (lower  left panel), and the  ${\rm 23^{\rm th}}$
        and ${\rm 24^{\rm th}}$ pulses of the $7 \, M_{\sun}$ computed
        with overshooting (lower right panel). In the figure are given
        the values  of $\Delta  M_{\rm CS}$, the  maximum mass  of the
        convective shell,  $\Delta M_{\rm H}$, the  mass through which
        the  hydrogen  profile moves  between  pulses, $\Delta  M_{\rm
        DU}$, the amount of  mass dredged--up during pulse power--down
        and $\Delta  M_{\rm EB}$,  the mass of  the outer edge  of the
        convective shell and the base of the convective envelope.  The
        mass  of the  hydrogen-exhausted  core is  shown  as a  dotted
        line.}
\label{dredge-up}
\end{figure*}

In Fig.~\ref{pulses}  we display the  helium luminosity for  the first
few thermal  pulses of some  selected models.  Note that  although the
time scales  are rather  different, the amplitudes  of the  pulses are
similar.  The fact that these  stars develop the TP--(S)AGB phase in a
way similar to that of solar metallicity models of analogous masses is
related  to the dredge--up  processes that  follow the  development of
shell helium burning.  The associated  metal enrichment of the base of
the  hydrogen--rich  envelope, as  well  as  its  heating due  to  the
proximity to the  helium burning shell (HeBS), allow,  as we have just
explained, for a new onset of hydrogen--burning through the CNO cycle,
that occurs as  soon as the distance between the HeBS  and the base of
the hydrogen--rich  envelope is about $10^{-4}\,  M_{\sun}$. From this
time  on, hydrogen  and helium  burning alternate  themselves  as main
energy suppliers.   In the next  two subsections we give  some details
about  the thermal  pulses that  ensue  in stars  that have  developed
either a carbon--oxygen or an oxygen--neon core.

\begin{figure*}[t]
\vspace*{12.5cm}
\hspace*{-1.7cm}
\includegraphics{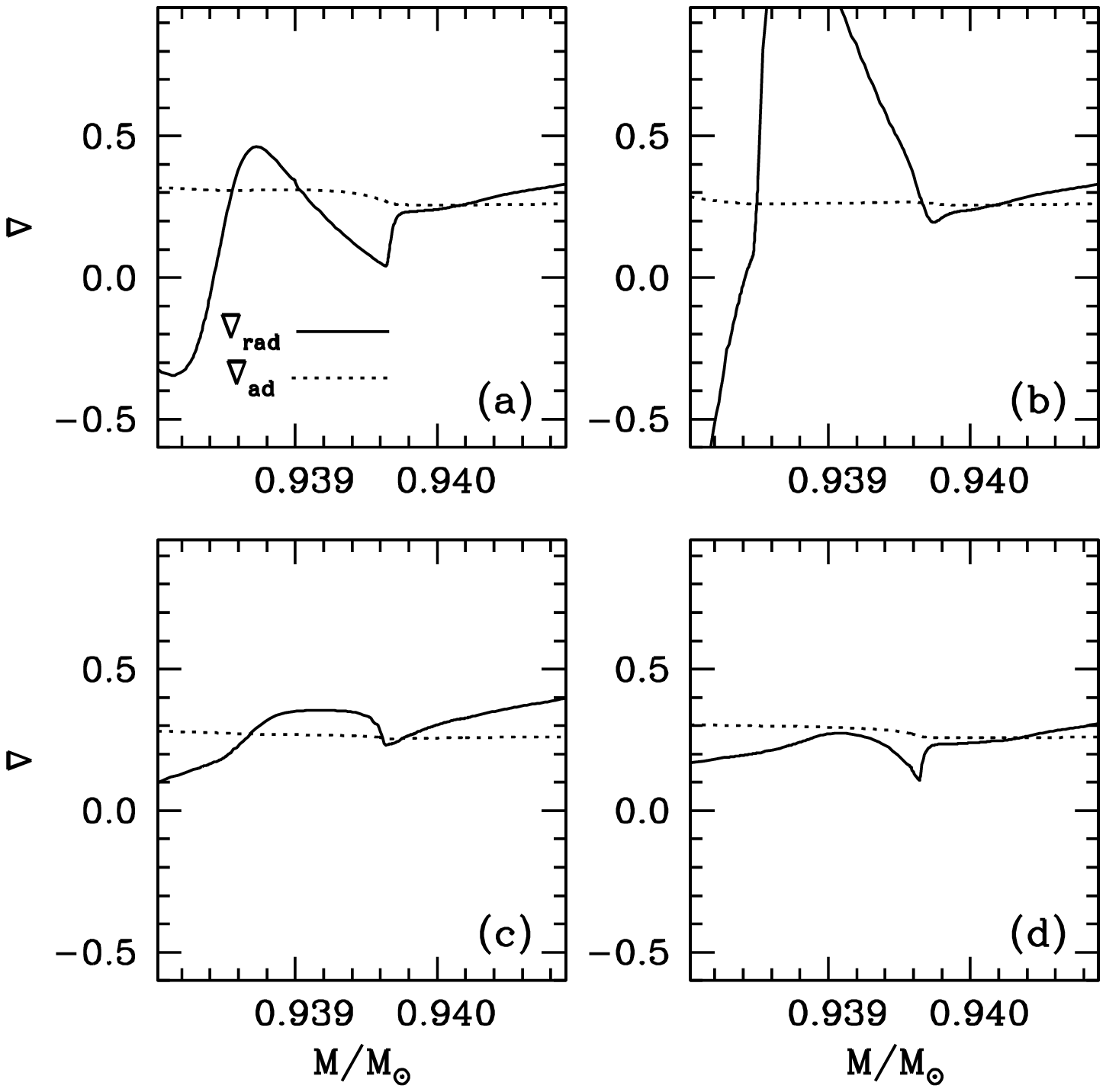}
\caption{Profiles  of the  adiabatic  gradient (dotted  line) and  the
         radiative  gradient  (solid line)  around  the $20^{\rm  th}$
         thermal  pulse  of  the  $5\,M_{\sun}$  model  computed  with
         overshooting. Panel  a shows  the situation near  the maximum
         luminosity    of    the    pulse    (at    $t=2.4341617\times
         10^{15}$~s).  Panel  b  depicts  the same  quantities  at  at
         $t=2.4341619\times  10^{15}$~s,   which  corresponds  to  the
         maximum extent  of the inner  convective zone. Panel  c shows
         the gradients near the maximum advance inwards of the BCE (at
         $t=2.4341645\times  10^{15}$ s). Finally,  panel d  shows the
         gradients  at the  end  of the  pulse (at  $t=2.4341699\times
         10^{15}$ s).}
\label{grads}
\end{figure*}

\begin{figure*}[t]
\vspace*{12.5cm}
\hspace*{-1.7cm}
\includegraphics{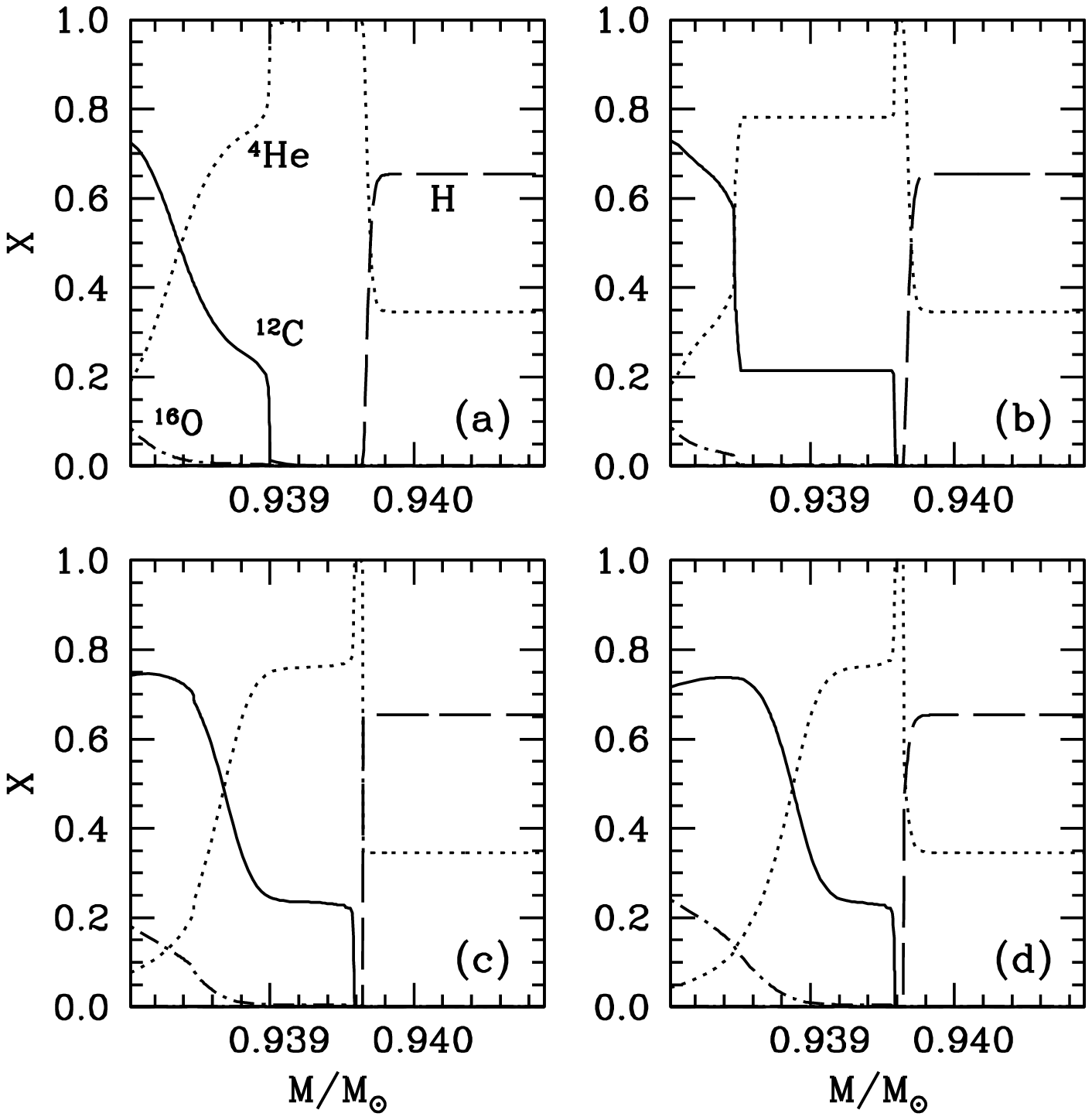}
\caption{Chemical profiles  of the main chemical species  for the $5\,
         M_{\sun}$ model star during  the $20^{\rm th}$ thermal pulse.
         The  evolutionary times are  those of  Fig.~\ref{grads}.  The
         profiles  are given at  the maximum  luminosity of  the pulse
         (panel a), at the maximum extent of the inner convective zone
         (panel b), near the maximum advance inwards of the BCE (panel
         c) and at the end of the pulse (panel d).}
\label{chemistry}
\end{figure*}

\subsection{Models with carbon--oxygen cores}

To keep  consistency with  the existing literature,  we will  refer to
these models  as TP--AGB  models. The upper  and lower left  panels of
Fig.~\ref{pulses} show, respectively, the  first few thermal pulses of
the  $7\,  M_{\sun}$ model  computed  without  overshooting and  those
corresponding  to   the  $5\,  M_{\sun}$  model   star  computed  with
overshooting.    These   are   representative  examples   of   massive
intermediate--mass  primordial stars  with  carbon--oxygen cores.   We
recall  that  models with  masses  at  the  ZAMS smaller  than  $7.8\,
M_{\sun}$ never  ignite carbon if no overshooting  is adopted, whereas
if overshooting  is taken  into account the  mass limit is  about $6\,
M_{\sun}$.  Since both models fail to ignite carbon, the TP--AGB phase
starts at the  end of the CHeB phase. However,  it is worth mentioning
that before reaching the TP--AGB phase, both models have experienced a
brief (and  weak) carbon--burning phase,  with associated luminosities
which never exceed  $\sim 10\, L_{\sun}$. This phase  is, in any case,
unable   to  modify   substantially  the   chemical  profile   of  the
corresponding  cores.  As  it  happens for  solar metallicity  TP--AGB
models,  each  helium  flash  is  accompanied  by  the  expansion  and
subsequent cooling of the  neighboring upper layers and, therefore, by
the switch--off of the  hydrogen burning shell.  The interpulse period
for the $7\, M_{\sun}$  model computed without overshooting amounts to
3200~yr, whereas it is reduced to 2700~yr for the $5\, M_{\sun}$ model
computed with overshooting.   As it can be seen  the average values of
the helium  luminosity are very  similar ($L_{\rm He} \approx  10^4 \,
L_{\sun}$)  for both  models,  but the  maximum  helium luminosity  is
somehow  larger   for  the  $7\,  M_{\sun}$   model  computed  without
overshooting.

The  upper  left  panel   of  figure  \ref{dredge-up}  represents  the
evolution of the BCE and the pulse--driven convective zones associated
to the  $8^{\rm th}$ and the  $9^{\rm th}$ helium flashes  of the $7\,
M_{\sun}$ computed without overshooting,  occurring at times $t \simeq
12.3446  \times 10^{14}$~s  and $t  \simeq 12.3455  \times 10^{14}$~s,
respectively.   It can  be  seen  that the  fast  injection of  energy
associated to  each helium  flash causes the  development of  an inner
convective zone, as  it occurs in the case  of solar metallicity model
stars.  Simultaneously,  the expansion and cooling that  occurs in the
neighboring  layers produce  the  extinction of  the hydrogen  burning
shell and,  as a consequence, a  fast advance inwards of  the BCE.  By
the  end of  the helium  flash,  the H--burning  shell moves  outward,
increasing the  mass of the He buffer  and a new He  flash occurs once
enough  matter is accumulated.   However, the  penetration of  the BCE
into the  regions that have  been processed by  previous pulse--driven
convective zones  is hardly noticeable. Hence, mixing  of the material
of the  hydrogen--exhausted core with that of  the convective envelope
does not allow  for a significant variation of  the metallicity of the
envelope and does  not affect the subsequent evolution.   In fact, the
increase in  metallicity that this model star  experiences during each
helium flash  is only $\Delta  Z_{\rm env} \approx 10^{-8}$.   For the
sake  of completeness,  we have  included in  Fig.~\ref{dredge-up} the
values of the maximum mass  of the convective shell during the $9^{\rm
th}$ pulse,  ($\Delta M_{\rm CS}=1.61\times  10^{-3}\, M_{\sun}$), the
mass  through  which the  BCE  moves  between  pulses ($\Delta  M_{\rm
H}=6.75\times  10^{-4}\, M_{\sun}$),  the mass  dredged--up  after the
$9^{\rm  th}$ helium  flash ($\Delta  M_{\rm  DU}=6.17\times 10^{-4}\,
M_{\sun}$), and the thickness of the region between the outermost edge
of  the convective  shell  and  the base  of  the convective  envelope
($\Delta M_{\rm EB}=1.12\times 10^{-4}\, M_{\sun}$).

To this regard it is important  to mention here that the efficiency of
the third  dredge--up strongly depends,  not only on the  treatment of
convection, but also on the numerical treatment.  Therefore, it is not
surprising  that   the  different  authors  that   have  studied  this
evolutionary  phase  have obtained  different  results. For  instance,
Marigo et al.  (2001) did not find any evidence of third dredge--up in
their calculations, although, as  these authors point out, this effect
could be due to the  small number of thermal pulses followed.  Chieffi
et  al.  (2001)  treated the  convective boundaries  according  to the
prescription  of Herwig  et al.   (1997) and,  moreover,  extended the
mixing  below  the  convective  envelope   in  such  a  way  that  the
composition   discontinuity   between   the   helium--rich   and   the
hydrogen--rich layers  was considerably smoothed.   Accordingly, these
authors found  a significant third  dredge--up after about  13 thermal
pulses. This  result is similar  to the one  obtained by Siess  et al.
(2002).   Unfortunately,   the  treatment  of   convection  should  be
validated by  calibrating theoretical models  with observations, which
is not possible  for primordial stars.  However, it  is worth pointing
out  that  even authors  whose  results  for  solar metallicity  stars
support  the existence of  an efficient  third dredge--up  (Karakas \&
Lattanzio 2002), have found some  evidences that might point to a less
efficient third dredge--up for  the case of metal--poor stars (Karakas
et al.  2006).   Finally, it is also important to  realize that in our
calculations we have not included the hot--bottom burning. In fact, it
can be  expected that hot--bottom burning would  increase the relative
abundance  in the  envelope of  nitrogen  and oxygen  with respect  to
carbon, but  not that it  changes critically the total  metallicity of
the envelope.   So, even  though we have  followed a  relatively small
number of  thermal pulses  (up to 20  at most),  we do not  expect any
drastic  effect  on  our  results   due  to  extra  pollution  by  hot
bottom--burning.
 
The  lower left  panel  of figure  7  represents the  behavior of  the
convective  regions  of  the  $5  \,  M_{\sun}$  model  computed  with
overshooting in  the period  of time elapsed  from shortly  before the
beginning of the $15^{\rm th}$  pulse until shortly after the $16^{\rm
th}$ pulse is  over. In this case, the  interpulse interval amounts to
2700 yr, somewhat shorter than that  of the $7 \, M_{\sun}$ model star
previously  discussed.  Moreover,  the evolution  of the  base  of the
hydrogen--rich  convective  envelope   and  of  the  convective  zones
associated to the helium flashes is totally equivalent to that already
done  and  the surface  composition  of  this  model does  not  change
appreciably. The  values of $\Delta  M_{\rm CS}$, $\Delta  M_{\rm H}$,
$\Delta M_{\rm DU}$ and $\Delta  M_{\rm EB}$ are also displayed in the
lower left panel of figure \ref{dredge-up}.

In  order  to  get  a  better  insight of  the  evolution  during  the
occurrence  of   a  helium  flash,  in  figure   \ref{grads}  we  have
represented as a dotted line the adiabatic gradient ({\rm}$\nabla_{\rm
ad}$)  and as a  solid line  the radiative  gradient ({\rm$\nabla_{\rm
rad}$}) in the region near the BCE, for some selected times during the
$20^{\rm  th}$ thermal pulse  of the  $5.0\, M_{\sun}$  model sequence
computed   with   overshooting.   Panel   (a)   corresponds  to   time
$t=2.4341617\times 10^{15}$~s.  At this  time the helium luminosity is
$L_{\rm He}\simeq  2.3 \times 10^4 L_{\sun}$ and  the convective shell
is  just  forming. Note  as  well that  for  this  model the  hydrogen
luminosity is rather small  $L_{\rm H}=4.4\, L_{\sun}$ and, therefore,
the BCE  is able to begin its  advance inwards. In the  zones in which
the radiative gradient is larger  than the adiabatic one, according to
the  (modified)  Schwarzschild criterion  used  for our  calculations,
there  is  complete  mixing  of  the  chemical  elements.   Panel  (b)
corresponds  to  time $t=2.4341619\times  10^{15}$  s,  for which  the
pulse--driven convective region reaches its maximum extension ($\Delta
M=1.36 \times 10^{-3}\, M_{\sun}$).  Panel (c) shows $\nabla_{\rm ad}$
and $\nabla_{\rm rad}$ at time $t=2.4341645\times 10^{15}$ s, when the
BCE reaches its maximum extent  inwards, at $M_r = 0.9397\, M_{\sun}$.
Finally, panel  (d) corresponds to time  $t=2.4341699\times 10^{15}$ s
and shows  the recession of  the BCE once  the HBS is  fully operative
again ($L_{\rm  H}=1.2\times 10^4\, L_{\sun}$).   As can be  seen, the
total amount of mass dredged--up is very small even if overshooting is
taken into account.
            
Figure \ref{chemistry} shows the chemical profiles of the main nuclear
species at the times already selected  for panels (a) to (d) of figure
\ref{grads}. In  panel (a)  the $^4$He buffer  formed due  to hydrogen
shell burning during  the previous interpulse period can  be well seen
between the  mass coordinates  $M_r = 0.93900  \,M_{\sun}$ and  $M_r =
0.93963  \,M_{\sun}$.  Later, at  time $t=2.4341619\times  10^{15}$ s,
corresponding to panel (b), the effects in the chemical profile of the
convective HeBS  can be  seen.  Note the  very small thickness  of the
region  in which  He has  not been  appreciably depleted.   This small
region lies between the two  convective regions: the one associated to
the  inner  He--burning  shell  and  that associated  to  the  H--rich
envelope.  In  panel (b)  it can be  seen that  the mixing due  to the
pulse--driven inner convective zone, allows a very small enrichment in
$^{12}$C and  $^{16}$O of the regions  just below the  BCE.  Panel (c)
shows  the maximum  advance inwards  of  the BCE,  that is  ultimately
halted  by  the steep  composition  profiles  that  correspond to  the
products of  the HBS. Finally, in  panel (d), one can  clearly see the
advance of the  carbon profile, as a result of  HeBS during the pulse,
as well  as the early recession  of the BCE.  In  summary, even though
overshooting  has been  taken  into account,  the  amount of  material
dredged--up from the core to the convective envelope is very small.

\begin{table*}
\centering
\caption{Surface abundances  (by mass) of carbon,  oxygen and nitrogen
         at  the  end of  our  calculations  for  the model  sequences
         computed   without  overshooting   (top  section)   and  with
         overshooting (bottom section).}
\begin{tabular}{cccccc}
\hline
\hline
$M_{\rm ZAMS}/M_{\sun}$ & $X$(C) & $X$(N) & $X$(O) & C:N:O  & $Z$  \\
\hline
 5 & $5.9\times 10^{-9}$  & $2.0\times 10^{-9}$  & $1.1\times 10^{-11}$ 
& $1 : 3.0\times 10^{-1} : 2.0\times 10^{-2}$ & $8.0\times 10^{-9}$ \\
 6 & $1.2\times 10^{-7}$  & $3.1\times 10^{-9}$  & $4.1\times 10^{-11}$ 
& $1 : 3.0\times 10^{-2} : 4.0\times 10^{-4}$ & $1.2\times 10^{-7}$ \\
 7 & $2.8\times 10^{-6}$  & $5.7\times 10^{-9}$  & $5.2\times 10^{-9}$  
& $1 : 3.0\times 10^{-3} : 2.0\times 10^{-3}$ & $2.8\times 10^{-6}$ \\
 8 & $9.0\times 10^{-5}$  & $7.4\times 10^{-7}$  & $1.3\times 10^{-6}$  
& $1 : 8.0\times 10^{-3} : 1.0\times 10^{-3}$ & $9.1\times 10^{-5}$ \\
 9 & $2.1\times 10^{-4}$  & $2.0\times 10^{-6}$  & $2.9\times 10^{-6}$  
& $1 : 1.0\times 10^{-2} : 1.5\times 10^{-2}$ & $2.1\times 10^{-4}$ \\
\hline
 5 & $3.0\times 10^{-7}$  & $2.0\times 10^{-8}$  & $9.4\times 10^{-11}$ 
& $1 : 7.0\times 10^{-2} : 3.0\times 10^{-4}$ & $3.2\times 10^{-7}$ \\
 6 & $3.3\times 10^{-5}$  & $3.8\times 10^{-7}$  & $1.6\times 10^{-7}$  
& $1 : 8.0\times 10^{-4} : 4.0\times 10^{-4}$ & $3.3\times 10^{-5}$ \\ 
 7 & $2.7\times 10^{-4}$  & $3.7\times 10^{-6}$  & $6.0\times 10^{-6}$  
& $1 : 1.0\times 10^{-2} : 2.0\times 10^{-2}$ & $2.7\times 10^{-4}$ \\ 
 8 & $2.8\times 10^{-3}$  & $1.4\times 10^{-4}$  & $4.7\times 10^{-6}$  
& $1 : 5.0\times 10^{-2} : 3.0\times 10^{-3}$ & $2.8\times 10^{-3}$ \\
\hline
\hline
\end{tabular}
\label{surface}
\end{table*}

\subsection{Models with oxygen--neon cores}

We  will refer  to models  developing oxygen--neon  cores  as TP--SAGB
stars.  The upper  and lower right panels of  figure \ref{pulses} show
the first thermal pulses of the $8 \, M_{\sun}$ model computed without
overshooting and the corresponding ones  for the $7 \, M_{\sun}$ model
star  computed   with  overshooting.   These   models  have  undergone
extensive   carbon   burning   and,   consequently,   have   developed
oxygen--neon cores.  At  the end of the carbon  burning phase the core
contracts while the helium burning shell is still active.  As the fuel
is exhausted in the innermost  layers, this shell advances outward and
approaches the  base of the hydrogen--rich  convective envelope.  When
the distance between  the latter and the HeBS is  of about $10^{-4} \,
M_{\sun}$ the base of  the hydrogen--rich convective envelope slightly
recedes and the small  radiative H--rich region is heated.  Eventually
the temperature  in this region  reaches the critical  temperature for
the onset  of the CNO--cycle  and the HBS is  reactivated.  Therefore,
the  TP--SAGB phase  develops in  a very  similar way  to that  of the
models with carbon--oxygen cores.  The details of the pulses of the $8
\, M_{\sun}$ model star computed without overshooting and of the $7 \,
M_{\sun}$ model in which overshooting was taken into account are shown
in the upper and lower right panels of figure \ref{dredge-up}.  Again,
our calculations  show that  the enrichment in  metals of  the stellar
surface is very small.

Table \ref{surface} lists the  surface abundances (by mass) of carbon,
nitrogen and  oxygen, their relative  abundance ratio (C:N:O)  and the
total  metallicity  ($Z$)  of  the model  sequences  computed  without
overshooting   (top   section   of   table  \ref{surface})   and   the
corresponding ones  when overshooting  was taken into  account (bottom
section).  As  it can  be seen, the  total metallicity of  the surface
increases with increasing $M_{\rm ZAMS}$. The reason for this behavior
is that more massive models have a lower degree of degeneracy in their
cores and, therefore,  the base of the convective  envelope is able to
penetrate  deeper  into  the   star  and,  consequently,  is  able  to
dredge--up larger amounts of material  which has been processed by the
different nuclear burning shells.

Another important feature of the surface abundances presented in table
\ref{surface} is  that all models are carbon--rich,  regardless of the
adopted  criterion for  convection.  For  the models  computed without
overshooting  (top  section   of  table  \ref{surface})  the  relative
abundance of nitrogen with  respect to carbon first decreases, reaches
a  minimum at  $M_{\rm ZAMS}\simeq  7\, M_{\sun}$  and  then increases
again. The same happens for the models computed with overshooting, but
in  this  case the  minimum  is  located  at $M_{\rm  ZAMS}\simeq  6\,
M_{\sun}$.   A similar behavior  is obtained  for the  relative oxygen
abundance  with  respect to  carbon  in  the  models computed  without
overshooting. For the  case in which overshooting has  been taken into
account the  relative oxygen abundance shows the  reverse behavior: it
first increases up  to a maximum at $M_{\rm  ZAMS}\simeq 7\, M_{\sun}$
and  then decreases. The  relative abundances  of nitrogen  and oxygen
with respect  to carbon, can be  better understood if  the models that
develop   carbon--oxygen   or   oxygen--neon  cores   are   considered
separately.   For  instance, the  abundances  of  nitrogen and  oxygen
relative to carbon decrease as the $M_{\rm ZAMS}$ increases for models
with a carbon--oxygen core, and  increase for the models which develop
an oxygen--neon core.


\section{The properties of the cores}

\begin{figure}[t]
\vspace{7.8cm}
\hspace{-2.7cm}
\includegraphics{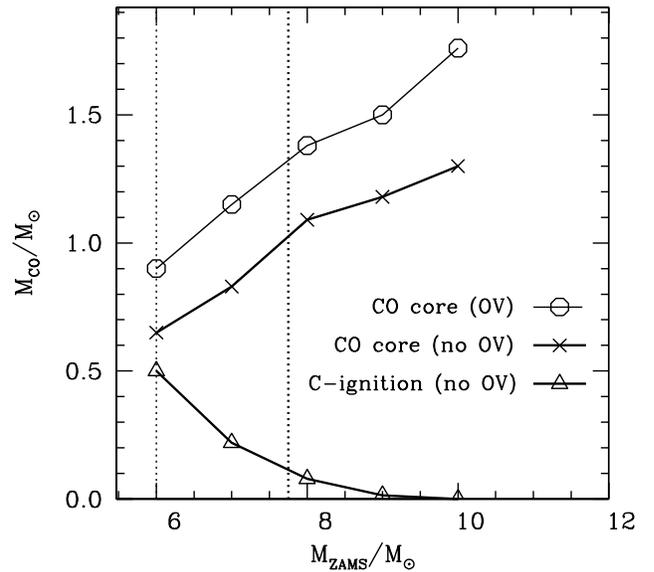}
\caption{Masses  of  the carbon--oxygen  cores  after  the central  He
         burning  phase for the  models computed  without overshooting
         (crosses and thick line)  and with overshooting (open circles
         and  thin  line).   The  mass  coordinate  for  which  carbon
         ignition  occurs is  also displayed  for the  models computed
         without  overshooting (triangles  and thick  line).  Finally
         the mass thresholds at the ZAMS for carbon burning to develop
         extensively are shown as vertical dotted lines.}
\label{sizes}
\end{figure}

Fig.~\ref{sizes} shows  the mass of the carbon--oxygen  core after the
central  He burning phase  for the  models computed  with overshooting
(open  circles and thin  line) and  without overshooting  (crosses and
thick  line)  as  a function  of  the  mass  at  the ZAMS.   The  mass
coordinate  where  carbon  ignition  initially occurs  is  also  shown
(triangles  and  thick  line)  for the  models  computed  disregarding
overshooting.    Carbon  ignition   for  the   models   computed  with
overshooting is  always central  for this mass  range.  The  ZAMS mass
corresponding to the intersection of the thick lines sets a lower mass
limit for single $Z=0$ stars to develop ONe degenerate cores. However,
our calculations show  that even though this is  an actual lower limit
to the mass threshold for the  onset of carbon ignition, the real mass
threshold for the  formation of an ONe degenerate core  is about $2 \,
M_{\sun}$ larger, as stars of  masses at the ZAMS between about $5.5\,
M_{\sun}$  and  $7.8  \,  M_{\sun}$  do  not  develop  carbon  burning
extensively  enough  in  the  carbon--oxygen  core  to  let  the  core
compositions  be  significantly changed.   The  dotted vertical  lines
correspond to these  critical masses below which carbon  is burnt only
partially.  For the case of the models computed with overshooting this
mass  threshold is  located  at $M_{\rm  ZAMS}\simeq 6.0\,  M_{\sun}$,
whereas  for the  models  in  which overshooting  was  not taken  into
account the mass threshold is $M_{\rm ZAMS}\simeq 7.8 \, M_{\sun}$.

We have followed the evolution of all our model stars until the end of
the carbon  burning phase  (if this is  the case) including  the early
stages of the thermally pulsing  AGB phase. Consequently, we have been
able  to determine the  chemical composition  of the  outer convective
envelope and of  the degenerate cores resulting from  the evolution of
massive  intermediate   mass  stars,  both  for  the   case  in  which
overshooting  was  considered  and  for  the  case  in  which  it  was
disregarded.  As  it will  be justified in  section \S 6  with further
detail ---  where we will  discuss the possible  evolutionary outcomes
for this  stellar mass  range ---  the final destiny  of the  cores of
these stars is not necessarily to form single white dwarfs.  This only
may occur if  a mechanism exists able to  produce extensive mass--loss
and, hence, able to remove  the convective envelope and leave behind a
compact remnant. Such  a mechanism could be, for  instance, Roche lobe
overflow  due to  the presence  of  a close  companion, or,  possibly,
stellar winds powered  by the (otherwise small) presence  of metals in
the envelope or, finally,  by chromospherically driven winds (van Loon
2005). Whatever  this mechanism could be  --- if it  indeed occurs ---
the  mass  and  the  chemical  composition of  the  resulting  compact
remnants is also a by--product  of our calculations.  We discuss these
properties in the following subsections.

\begin{figure}[t]
\vspace{7.9cm}
\hspace{-2.7cm}
\includegraphics{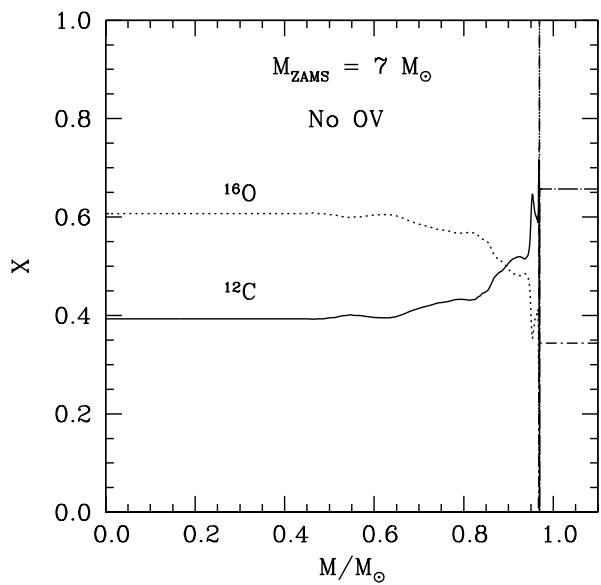}
\caption{Composition profile of the degenerate CO core for the $M_{\rm
         ZAMS}=7 \, M_{\sun}$ case without overshooting.}
\label{core7noOV}
\end{figure}
       
\subsection{Models computed without overshooting}

For the  case in  which no overshooting  is adopted,  primordial stars
with masses at  the ZAMS smaller than $7.8\,  M_{\sun}$ do not undergo
extensive  carbon burning.   The pollution  in metals  of  the stellar
envelope is  not large (see table  \ref{surface}).  Hence, radiatively
driven stellar  winds during  this stage are  not likely to  be strong
enough  to remove  the  outer  envelopes of  these  stars and  produce
carbon--oxygen  white  dwarfs, as  it  would  be  the case  for  solar
metallicity  stars of equivalent  mass.  Instead,  the cores  of these
stars, due to the successive  thermal pulses, could eventually grow up
to the Chandrasekhar mass ($M_{\rm Ch}$). Thus, these stars are likely
to undergo  a supernova explosion.   This possibility will  be further
explored and  quantified in  \S 6.   If this is  indeed the  case, the
chemical  composition of  the  remnant  core is  of  interest.  As  an
example of the resulting degenerate core, figure \ref{core7noOV} shows
the composition profile of the degenerate core resulting from a $7.0\,
M_{\sun}$, computed without overshooting.
                                                                         
\begin{figure}[t]
\vspace{7.9cm}
\hspace{-2.7cm}
\includegraphics{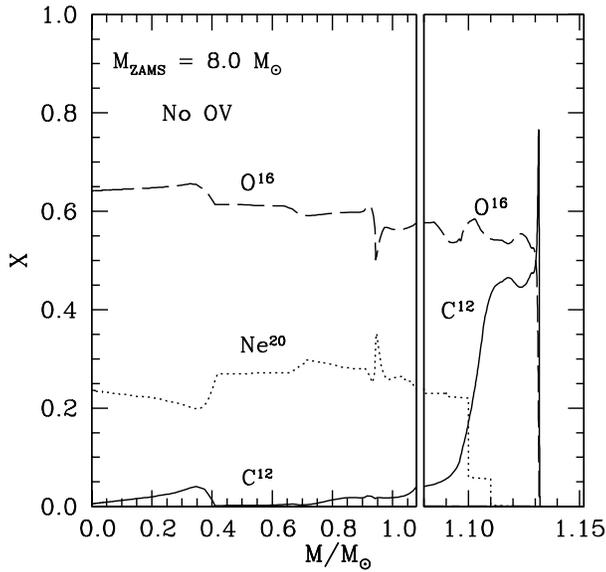}
\caption{Composition  profile  of  the  degenerate ONe  core  for  the
         $M_{\rm ZAMS}=8 \, M_{\sun}$ model without overshooting.}
\label{core8noOV}
\end{figure}

\begin{table*}
\centering
\caption{Characteristics  of  the remnant  cores  at  the  end of  our
         calculations  for   the  model  sequences   computed  without
         overshooting  (top  section)  and with  overshooting  (bottom
         section).}
\begin{tabular}{cccccc}
\hline
\hline
$M_{\rm ZAMS}/M_{\sun}$ & 
$M_{\rm CO}/M_{\sun}$ & 
$M_{\rm ONe}/M_{\sun}$ & 
$\Delta M_{\rm CO}/M_{\sun}$ & $X$(C)/$X$(O)& $X$(Ne)/$X$(O)\\
\hline
 5 &  0.90    &  ---   &   ---   & 0.60 & ---  \\
 6 &  0.92    &  ---   &   ---   & 0.62 & ---  \\
 7 &  0.96    &  ---   &   ---   & 0.65 & ---  \\
 8 &  ---     &  1.01  &   0.025 & ---  & 0.56 \\
 9 &  ---     &  1.18  &   0.015 & ---  & 0.47 \\
10 &  ---     &  1.35  &   0.010 & ---  & 0.35 \\
\hline
 5 &  0.94    &  ---   &   ---   & 0.60 & ---  \\
 6 &  ---     &  1.00  &   0.070 & ---  & 0.50 \\
 7 &  ---     &  1.20  &   0.200 & ---  & 0.42 \\
 8 &  ---     &  1.34  &   0.002 & ---  & 0.36 \\
 9 &  ---     &  1.46  &   0.008 & ---  & 0.35 \\
\hline
\hline
\end{tabular}
\label{cores}
\end{table*}

According to  our calculations,  $Z =  0$ stars in  the range  $7.8 \,
M_{\sun}  \la  M_{\rm  ZAMS}  \la 9.0  \,M_{\sun}$  experience  carbon
burning  in conditions  of partial  degeneracy if  no  overshooting is
adopted.  Thus,  these stars form  oxygen--neon cores surrounded  by a
carbon--oxygen degenerate  buffer.  For this  mass range we  also find
that the  enrichment in  metals of the  convective envelope  is rather
small  and, hence,  are prone  to undergo  accretion  induced collapse
(AIC) leading also to  a supernova explosion. The chemical composition
of  such cores is  also relevant  for the  particulars of  their final
evolutionary phases --- see,  for instance, Guti\'errez et al.  (1996)
and  Guti\'errez  et al.   (2005).   Fig.   \ref{core8noOV} shows  the
internal  chemical  profile  for  the  case  of  a  $M_{\rm  ZAMS}=8\,
M_{\sun}$  primordial  star.   Finally,  for  the  case  in  which  no
overshooting is adopted, primordial  stars of masses $M_{\rm ZAMS} \ga
9.5\, M_{\sun}$ undergo extensive carbon burning if no overshooting is
taken  into account,  but the  resulting oxygen--neon  cores  are only
weakly  degenerate.   Such  stars   would  most  likely  become  SNeII
directly.

The  top section  of table  \ref{cores} shows  a summary  of  all this
information. In particular, the mass of the CO core (second column) or
the ONe  core (third  column) --- depending  on whether the  model has
avoided carbon ignition  or not --- at the  beginning of the thermally
pulsing  phase  are displayed.   Also  shown  are  the masses  of  the
remaining CO buffers on top of the ONe cores ($\Delta M_{\rm CO}$) for
the models which have been  able to ignite carbon (fourth column), the
carbon--to--oxygen ratio in  the CO cores of the  models which fail to
ignite carbon  (fifth column) and  the neon--to--oxygen ratio  for the
models which ignite  carbon (last column).  As can  be seen, for these
evolutionary sequences the mass of the carbon--oxygen buffer decreases
as the mass of the ONe core increases and so does the neon--to--oxygen
ratio.

\subsection{Models computed with overshooting}

\begin{figure}[t]
\vspace{7.9cm}
\hspace{-2.7cm}
\includegraphics{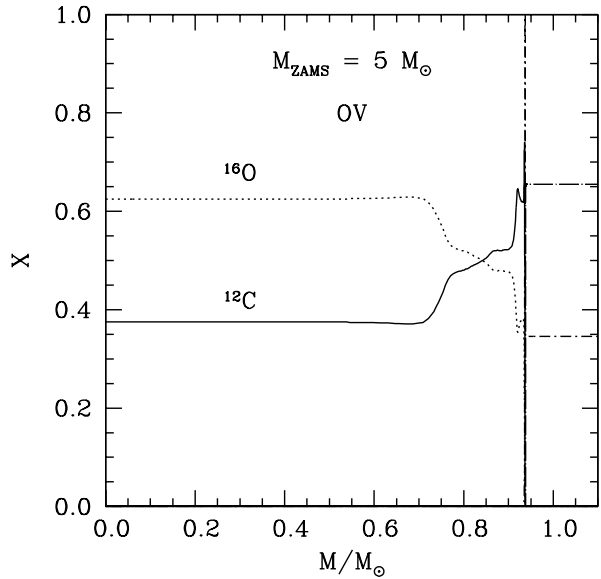}
\caption{Composition profile of the degenerate CO core for the $M_{\rm
         ZAMS}=5 \, M_{\sun}$ case with overshooting.}
\label{core5OV}
\end{figure}
                     
The minimum  mass at  the ZAMS  of a primordial  star that  allows for
extensive  carbon  ignition  and,  hence,  for  the  formation  of  an
oxygen--neon core is, as  previously mentioned, $6.0\, M_{\sun}$.  The
convective  envelopes of  stars beyond  this mass  threshold  are more
metal rich  than those  computed without overshooting.   Therefore, we
expect  that  these  stars   could  potentially  be  able  to  support
radiatively  driven winds and  loose their  envelopes, leading  to the
formation  of carbon--oxygen  white  dwarfs.  In  this  case the  core
composition of  these stars  is also of  interest because  the cooling
speed depends on the carbon  to oxygen ratio (Segretain et al.  1994).
Figure \ref{core5OV} shows an  example of the resulting carbon--oxygen
core for the $M_{\rm ZAMS}=5\, M_{\sun}$ model sequence.

\begin{figure}[t]
\vspace{7.9cm}
\hspace{-2.7cm}
\includegraphics{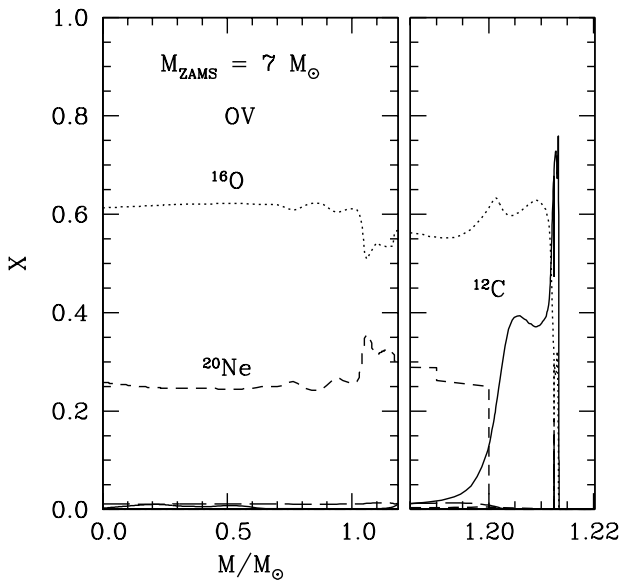}
\caption{Composition  profile  of  the  degenerate ONe  core  for  the
         $M_{\rm ZAMS}=7 \, M_{\sun}$ model with overshooting.}
\label{ONe7OV}
\end{figure}
                  
\begin{figure}[t]
\vspace{7.9cm}
\hspace{-2.7cm}
\includegraphics{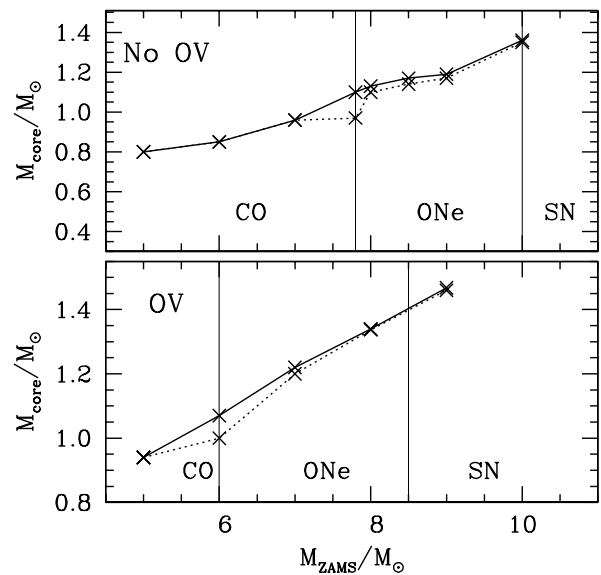}
\caption{Mass  of  the  compact  remnant  cores  by  the  end  of  our
        calculations for the cases  with no overshooting (upper panel)
        and with overshooting (lower panel).  The solid line shows the
        total mass  of the degenerate  core, while the dotted  line is
        the mass of the ONe  core --- when applicable --- without the
        CO buffer.}
\label{coregraph}
\end{figure}

Primordial stars with masses at the ZAMS between $6.5 \, M_{\sun}$ and
$8.5\, M_{\sun}$  do ignite  carbon at the  center and  may eventually
become  oxygen--neon  white   dwarfs,  provided  that  the  convective
envelope  is removed by  some mechanism  during the  thermally pulsing
phase. We  note however  that in  this case the  metal content  of the
convective envelope  is sizeable --- see  table 2 ---  and, thus, this
stars  are prone  to loose  their convective  envelopes as  well. Fig.
\ref{ONe7OV}  shows the  typical final  chemical profiles  of  such an
oxygen--neon white dwarf, which do not differ substantially from those
of stars of solar metallicity.   Finally, for stars with masses at the
ZAMS larger than $8.5\, M_{\sun}$  the hydrogen burning shell does not
resurrect  and, thus,  these stars  do  not have  a thermally  pulsing
phase.  Moreover, the mass of  their oxygen--neon cores is larger than
$M_{\rm Ch}$ and, consequently,  their evolution continues all the way
to  more advanced  stages  of  stellar evolution  and  they will  most
probably produce  a supernova explosion.  The bottom  section of table
\ref{cores}  shows a  summary  of all  this  information. Finally,  in
figure \ref{coregraph} we present all this information graphically for
both  the case  in which  overshooting  is adopted  (bottom panel)  or
neglected (upper panel). In both  panels the solid lines represent the
total mass  of the degenerate cores  and the dotted lines  the mass of
the ONe  cores.  We  have also separated  and labeled the  regions for
which at the end of our calculations  we obtain a CO core, an ONe core
or  directly a  supernova. Note,  however, that  the final  outcome of
these stars still remains to be studied.  This will be the goal of our
next section.

\section{Mass loss and the final fate of primordial massive 
         intermediate--mass stars}

In this section  we will consider how the  uncertainties that surround
the evolution of primordial TP--(S)AGB stars can affect their expected
final  fate.   We  will  show  that, taking  into  account  reasonable
prescriptions for the mass loss rates during the TP--(S)AGB phase, the
possibility that  some of  these stars end  their lives as  SNI1/2 ---
instead of  white dwarfs, as it  is usually assumed  --- remains open.
These supernovae  are produced by  the explosion of  a CO core  in AGB
stars (Arnett 1969,  Nomoto 1976).  We will also  show that primordial
stars that  form ONe cores  near the Chandrasekhar mass  will possibly
undergo an electron--capture induced  collapse to a neutron star, very
likely producing a  type II--P supernova. Our study  can be considered
as  a first  approximation to  the problem,  since following  the full
evolution of these  kind of stars until the very  late phases of their
evolution  is clearly  out  of our  current possibilities.   Synthetic
(S)AGB models  like those of  Izzard et al.   (2004) can help  in this
regard.  Such an study  is currently  under way,  and we  postpone the
discussion of such synthetic models to a forthcoming publication.

The particular details of the  evolution of stars of intermediate mass
from the end of the AGB phase, through a planetary nebula phase, until
the formation  of a white dwarf  are not entirely  well understood for
solar metallicity stars. However, there  is a general agreement on the
overall  picture. Strong  winds are  expected  to remove  most of  the
stellar envelopes  before any significant  growth of the cores  due to
shell  burning processes  is achieved.   Therefore, the masses  of the
remnant cores  after the  main central burning  phases provide  a good
approximation to the final mass of the white dwarfs resulting from the
evolution of these stars.

However, the  uncertainties associated to the mass--loss  rates due to
stellar winds  affect the details  of the late evolutionary  stages of
solar  metallicity stars.   For the  case of  primordial  stars, these
uncertainties are  more dramatic because,  as it will be  shown below,
the  adopted mass--loss  rates can  change  the final  outcome of  the
stellar evolutionary sequences and let  them end their lives either as
a white dwarf  or as a supernova.  In  principle, the mass--loss rates
are expected to be smaller for stars of low metallicity. For instance,
a common prescription  to take this fact into account  is to adopt the
following expression:

\begin{equation}
\dot M(Z) = \dot M(Z_{\sun})\;\left(\frac{Z}{Z_{\sun}}\right)^{0.5}
\end{equation}

\noindent where  $\dot M(Z_{\sun})$ is  the mass--loss rate  for solar
metallicity.  Note,  however, that  this prescription is  still rather
uncertain  and that  it is  the subject  of a  vivid debate  (van Loon
2005). However, following closely  the recent work of Ziljstra (2004),
which  is suggestive  of a  break--down  of the  mass--loss rates  for
dust-driven winds  when $Z<0.1\,Z_{\sun}$, we adopt it  as an educated
guess.  Other  recent works  that have used  this expression  to scale
mass loss with metallicity are Kudritzki (2000), Vanbeveren (2001) and
Woosley et al. (2002).

There are  several prescriptions in the literature  that provide $\dot
M(Z_{\sun})$.  One may use, for instance, the expression of Schr\"oder
\& Cuntz (2005),  whose authors claim it to  yield reasonable results,
even for the tip of the AGB:

\begin{eqnarray}
\dot M_{\rm SC}(Z_{\sun}) &=& 
\dot M_{\rm R}(Z_{\sun}) \left({\frac {T_{\rm eff}}{4000}}\right)^{3.5}\\
&{\;}&\left(1 + \frac{g_{\sun}}{4300\;g}\right)\;\nonumber
{M_{\sun}}\,{\rm{yr}^{-1}}
\end{eqnarray}

\noindent  where $T_{\rm  eff}$ is  the effective  temperature  of the
star, $g$ is its surface gravity, $g_{\sun}$ is the surface gravity of
the Sun and $\dot M_{\rm R}$ is the mass loss rate of Reimers (1975):

\begin{equation}
\dot M_{\rm R}(Z_{\sun}) = -4\times 10^{-13} \eta_{\rm R}\; \frac{L\,R}{M}\;
M_{\sun} \,\rm{yr}^{-1}
\end{equation}

\noindent  where  $\eta_{\rm R}$  is  a  parameter  such that  $1/3  <
\eta_{\rm  R} <  3$.  Another  common prescription  is  the expression
proposed by Bl\"ocker (1995):

\begin{equation}
\dot M_{\rm B}(Z_{\sun}) = -4.83\times 10^{-9} \frac{L^{2.7}}
{{M_{\rm TP}}^{2.1}}
\dot M_{\rm R}(Z_{\sun}) \; M_{\sun} \,\rm{yr}^{-1}
\end{equation}

\noindent where $M_{\rm TP}$ is the actual mass of the considered star
and, therefore, decreases as mass is lost. For solar metallicity stars
this  prescription  yields   consistently  larger  values  during  the
TP--(S)AGB phase  (Willson 2000, Gallart  et al. 2005) as  compared to
the  rest of widely--used  prescriptions found  in the  literature. We
adopt  it  because  it  provides  an  upper  bound  for  the  expected
mass--loss rate.

\begin{figure}[t]
\vspace{7.9cm}
\hspace{-2.7cm}
\includegraphics{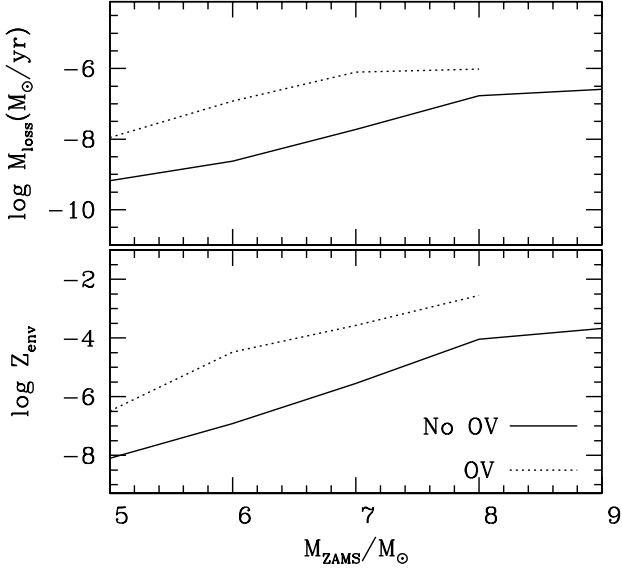}
\caption{Bottom  panel: metallicity  of the  envelope of  the selected
         models at  the end of our calculations.   Top panel: expected
         mass--loss  rates  given the  envelope  metallicities of  the
         bottom panel  when the Schr\"oder \&  Cuntz (2005) mass--loss
         rates are adopted.   Models computed without overshooting are
         represented  using  solid   lines  whereas  models  in  which
         overshooting was adopted are shown with dotted lines.}
\label{Z}
\end{figure}

When helium burning  sets in a shell and  during the thermally pulsing
phase  there   is  some   dredge--up  of  metal--rich   material  and,
consequently,  the originally  almost  primordial convective  envelope
becomes  polluted  in  metals.   Naturally,  the  metallicity  of  the
convective   envelope  depends   on  the   adopted   prescription  for
convection, and, of course, on the evolutionary model, but it is never
larger than  $Z\sim 2.8\times 10^{-3}$.  Taking  into account Eq.~(2),
with  the  different prescriptions  for  $\dot  M(Z_{\sun})$ given  by
Eq.~(3), with $\eta_{\rm  R}=1.0$, or by Eq.~(5), a  rough estimate of
the minimum  time required to  loose the envelope  through radiatively
driven stellar winds can be obtained:

\begin{equation}
t_{\rm env} = \frac{\Delta M_{\rm env}}{\dot M}
\end{equation}

\noindent where  $\Delta M_{\rm  env}$ is the  mass of  the convective
envelope. In  the bottom panel  of figure \ref{Z} we  have represented
the logarithm of the metallicity  of the envelope, $\log Z_{\rm env}$,
at the end of our calculations for  the 5, 6, 7, 8 and $9 \, M_{\sun}$
stellar  evolutionary sequences  computed without  overshooting (solid
line), and for the 5, 6, 7 and $8\, M_{\sun}$ model sequences computed
with overshooting (dotted  line).  For the case of  the $9\, M_{\sun}$
model  star computed with  overshooting we  did not  obtain dredge--up
and,  thus, the  envelope was  not significantly  enriched  in metals,
leading to $Z_{\rm  env}=10^{-11}$.  Additionally, the degenerate core
of this  model star has  a mass very  close to the  Chandrasekhar mass
and, thus,  the most likely outcome  is a supernova. The  top panel of
figure  \ref{Z}  represents  the  corresponding mass--loss  rates,  as
obtained by  adopting the prescription of Schr\"oder  \& Cuntz (2005).
We remark that these mass--loss rates could be taken as a lower limit,
as they have been computed  taking into account the metallicity of the
envelope at the  end of our calculations.  If the  position of the BCE
is,  approximately, sustained  and  the envelope  mass decreases,  the
freshly synthesized metals dredged--up  from the core will appear less
diluted in the stellar surface and hence will allow a certain increase
in  the  mass--loss  rates.  On  the  other hand,  the  mass  loss  of
primordial  stars   is  actually  hampered  by   the  compactness  and
relatively high effective temperatures of these stars that result as a
consequence of their low envelope metallicities. Since these stars are
more  compact the  material lost  by these  stars must  escape  from a
deeper  gravitational  well.    Additionally,  the  high  photospheric
temperatures  do  not  allow  for  the  formation  of  dust,  even  if
refractory elements are present.   Both facts will result into smaller
mass--loss rates.

Our evolutionary  models yield  the rate of  growth of  the degenerate
core, $\dot M_{\rm  core}$, as a consequence of  the successive thermal
pulses  and  also   provide  us  with  the  mass   of  the  degenerate
core. Hence, the time needed to reach the Chandrasekhar mass would be:

\begin{equation}
t_{\rm core} \sim \frac{M_{\rm Ch}- M_{\rm core}}{\dot M_{\rm core}}
\end{equation}

For  instance,  for  the  $7\,  M_{\sun}$  model  star  computed  with
overshooting  the  growth  rate  of  the degenerate  core  during  the
thermally pulsing  phase is $\sim 2.0 \times  10^{-7} \, M_{\sun}$/yr.
Since this model star has already an oxygen--neon core of mass $M_{\rm
ONe}\simeq 1.2  \, M_{\sun}$  --- see table  \ref{cores} ---  it would
only need about $1.0 \times  10^6$~yr to reach the Chandrasekhar mass.
Given the mass--loss rate shown in the top panel of figure \ref{Z} ---
about  $1.7\times 10^{-8} \,  M_{\sun}\; {\rm  yr}^{-1}$ ---  the time
needed to get rid of the  envelope would be about $3.5 \times 10^8$~yr
if the prescription  of Schr\"oder \& Cuntz (2005)  is adopted for the
mass--loss rate.   That is,  $t_{\rm env}$ is  about 350  times larger
than the  time needed by the  star to reach  $M_{\rm Ch}$.  Therefore,
one could  reasonably expect that  this star would  end its life  as a
supernova.

In  this respect,  it  must be  said that  our  work is  just a  first
approach, as we have assumed that the rate of core growth is constant,
that the  third dredge--up is  going to remain inefficient  during the
whole TP--(S)AGB phase,  and that we have not  included the effects of
hot-bottom burning.   A more realistic treatment of  the problem would
require to build up a synthetic code, as it has been done by Izzard et
al. (2004), or  Groenewegen \& de Jong (1994)  and references therein.
However,  there  are some  facts  that  support  the interest  of  our
results.   First of  all, the  fact that  the times  required  for the
growth of the core up to  $M_{\rm Ch}$ are several orders of magnitude
longer than  the times required to  get rid of  the envelopes. Second,
the fact  that even  when disregarding the  effect of  metallicity, we
still obtain  the same qualitative  behavior, although the  values of
the times required to eject the envelope decrease considerably. Third,
that using  other mass loss  prescriptions, such as Reimers  (1975) or
Vassiliadis \& Wood (1993), we also reproduce the same behavior.

The procedure explained above can be repeated for all the evolutionary
sequences  presented   here  adopting  either   the  prescriptions  of
Schr\"oder \& Cuntz (2005) and  of Bl\"ocker (1995) for the mass--loss
rates.  The results can be found in table \ref{timescales}.  The first
column  of  this table  displays  the  initial  masses of  the  models
analyzed in this work, the second column shows the time needed for the
core of these  stars to reach the Chandrasekhar  mass, $t_{\rm core}$,
as obtained  from Eq.~(7),  and using the  core growth  rates obtained
from our  evolutionary calculations.   In particular, the  core growth
rates  have been  obtained by  dividing the  mass growth  of  the core
between the  fifth and the  last pulse computed  in each model  by the
corresponding  time interval.  We  have avoided  the first  pulses (or
mini--pulses) in order to obtain  average core growth rates during the
whole  thermally pulsing  phase. With  this procedure  we  obtain core
growth rates  ranging from $2\times 10^{-7}  M_{\sun}\; {\rm yr}^{-1}$
to  $5\times 10^{-7} \,  M_{\sun}\; {\rm  yr}^{-1}$.  Because  we have
followed a limited number of pulses, these values could not correspond
exactly to  the average core  growth rates during the  whole thermally
pulsing phase but provide us  with a reasonable good approximation for
the purposes  of the present work. The  time needed to get  rid of the
envelope according to the  prescription of Schr\"oder \& Cuntz (2005),
$t_{\rm  env}^{\rm  SC}$,  is  shown  in the  third  column  of  table
\ref{timescales},  whereas  the  time  needed to  loose  the  envelope
according to  the prescription of Bl\"ocker  (1995), $t_{\rm env}^{\rm
B}$, is shown in  the fourth column.  As it has been  done so far, the
top section of table \ref{timescales} shows the results for the models
computed  without   considering  overshooting,  whereas   the  results
obtained  taking overshooting  into account  are shown  in  the bottom
section of this table.

\begin{table}
\centering
\caption{Timescales (in  years) associated to core growth  and loss of
         the envelope for the case  in which no overshooting was taken
         into  account  (top  section)  and  for  the  case  in  which
         overshooting was adopted (bottom section).}
\begin{tabular}{cccc}
\hline
\hline
$M_{\rm ZAMS}/M_{\sun}$ & $t_{\rm core}$ 
& $t_{\rm env}^{\rm SC}$ & $t_{\rm env}^{\rm B}$ \\
\hline
5 &  $3.0\times 10^6$ & $6.2\times 10^{9}$ & $2.6\times 10^{8}$ \\
6 &  $1.9\times 10^6$ & $2.1\times 10^{9}$ & $1.2\times 10^{8}$ \\
7 &  $1.2\times 10^6$ & $3.1\times 10^8$    & $7.7\times 10^{6}$ \\
8 &  $1.1\times 10^6$ & $4.1\times 10^7$    & $6.7\times 10^{5}$ \\
9 &  $0.7\times 10^6$ & $2.9\times 10^7$    & $3.4\times 10^{5}$ \\
\hline
5 &  $9.0\times 10^6$ & $4.1\times 10^8$ & $6.6\times 10^{6}$ \\
6 &  $7.2\times 10^5$ & $4.3\times 10^7$ & $7.0\times 10^{5}$ \\
7 &  $5.0\times 10^5$ & $6.8\times 10^6$ & $4.2\times 10^{4}$ \\
8 &  $1.7\times 10^5$ & $1.2\times 10^7$ & $3.1\times 10^{5}$ \\
\hline
\hline
\end{tabular}
\label{timescales}
\end{table}

If  we compare  the  times required  for  these stars  to loose  their
envelopes according to the  prescription of Schr\"oder \& Cuntz (2005)
with those required by the cores to reach $M_{\rm Ch}$, it can be seen
that for  all the models  in table \ref{timescales}  $t_{\rm env}^{\rm
SC}$ is between one and  three orders of magnitude larger than $t_{\rm
core}$.  Therefore, all these stars would likely end up their lives as
supernovae, unless some  additional mechanism enhancing the mass--loss
rate played a significant role in their evolution.  Rotational induced
mixing could be this extra mixing mechanism required for the existence
of  strong   stellar  winds.   Being  completely   devoid  of  metals,
primordial stars  would have  no means of  slowing down  their initial
rotation (Chiappini et al., 2006).  Furthermore, because they are more
compact  than  higher  metallicity  stars of  similar  masses,  higher
rotational  velocities could be  expected simply  as a  consequence of
angular  momentum conservation during  the initial  collapsing stages.
Hence  primordial  stars  might  experience a  substantial  amount  of
rotational mixing (Meynet et al., 2006). It is important to point that
overshooting  cannot  account for  this  extra  mixing, as  rotational
mixing  proceeds  also  in  radiative  zones.  In  particular,  it  is
important  to  note that  Meynet  et  al.   (2006) have  followed  the
evolution of a $7 \,  M_{\sun}$ star, with $Z=10^{-5}$, and have found
a 1000--fold increase in  its surface metallicity. Nevertheless, it is
also worth  noticing that the initial rotation  velocity considered in
this  work was  near the  critical  rotation velocity.   In any  case,
rotation is  always difficult to introduce in  a self--consistent way.
Some groups assume that zero  metallicity stars are fast rotators, but
other advocate, on the basis  of theoretical grounds, for much smaller
rotation rates  for primordial stars (Silk \&  Langer 2006). Moreover,
Yoon \&  Langer (2005)  point to another  open question:  the feedback
between rotation  and stellar winds.  These authors  conclude that the
rotation rates required  to allow for the mixing  of the collapsar can
only  be  kept  for  weak  stellar winds.   Consequently,  this  point
continues to be an open issue.

The picture  changes considerably when we consider  the times required
to loose  the envelope  when the prescription  of Bl\"ocker  (1995) is
adopted.   Not  surprisingly,   the  mass--loss  rates  increase  and,
therefore, the times required to loose the stellar envelopes decrease.
The  results  are the  following:  the  most  massive models  computed
without overshooting (the 8 and  $9 \, M_{\sun}$ model stars), and all
the models computed with overshooting except the $8 \, M_{\sun}$ model
are able  to loose  their envelopes before  their cores  reach $M_{\rm
Ch}$  and,  therefore,  would   probably  end  their  lives  as  white
dwarfs. The  $8 \, M_{\sun}$ model computed  with overshooting becomes
almost immediately a supernova, because the mass of its core ($1.34 \,
M_{\sun}$) is already very close to $M_{\rm Ch}$.  The key factor that
determines the long timescales required  to loose the envelope for the
models  with masses  between 5  and $7  \, M_{\sun}$  computed without
overshooting  is  the low  metallicities  of  the  envelopes of  these
models.    A  comparison   of   the  envelope   timescales  with   the
corresponding  values of  $t_{\rm core}$  allows us  to  conclude that
these objects are likely to end their lives as SNeI1/2, as it has been
previously found for all the cases computed adopting the Schr\"oder \&
Cuntz (2005) prescription for the mass--loss rates.

\section{Summary and Discussion}

We   have  followed  the   evolution  of   massive  intermediate--mass
primordial stars from the zero age  main sequence until the end of the
carbon burning phase and  the thermally pulsing (Super)--AGB phase (if
this is the  case).  We have also studied the  role of overshooting by
computing two series of evolutionary sequences.  In the first of these
series  overshooting was  completely  neglected.  For  this series  of
evolutionary  models we  followed the  evolution of  model  stars with
masses  ranging from  $5\, M_{\sun}$  to $9\,  M_{\sun}$. In  a second
series of  calculations overshooting was taken  into account according
to the  prescription of  Eldridge \& Tout  (2004) and we  followed the
evolution of primordial stars  with masses ranging from $5\, M_{\sun}$
to $8\, M_{\sun}$.  This has allowed us to determine the masses of the
degenerate cores  resulting after the  main central burning  stages of
primordial stars  of these  masses, and to  study the mass  limits for
primordial  stars to  form CO  white dwarfs,  ONe white  dwarfs  or to
undergo a supernova explosion.

Our main results can be summarized as follows: if we consider the case
in which overshooting was disregarded  we have obtained a maximum mass
for the  formation of CO degenerate  cores of $\sim  7.8 \, M_{\sun}$,
and a maximum mass  for the formation of an ONe cores  of $\sim 9.5 \,
M_{\sun}$.    Stars    more   massive   than    $9.5\,M_{\sun}$   form
electron--degenerate cores  with masses larger  than the Chandrasekhar
limit  and, consequently, would  directly undergo  supernova explosion
through  the  core  collapse  mechanism.   If  a  moderate  amount  of
overshooting  is adopted  the mass  threshold for  the formation  of a
carbon--oxygen  core, a neon--oxygen  core or  to undergo  a supernova
explosion  is shifted  to  smaller  masses by  $\sim  2 \,  M_{\sun}$.
Therefore, we  get a maximum  ZAMS mass of  $6.0 \, M_{\sun}$  for the
formation  of a  CO degenerate  core and  a maximum  value of  $8.5 \,
M_{\sun}$  for the  formation of  an  ONe degenerate  core. One  would
obtain a supernova  for ZAMS masses larger than  this value.  For both
sequences  of  models (those  in  which  overshooting  was taken  into
account and those in which it was disregarded) the resulting ONe cores
appear to  have a chemical  stratification similar to  that previously
found  by  Garc\'{\i}a--Berro et  al.   (1997)  and  Gil--Pons et  al.
(2005).   Namely, an  ONe  electron--degenerate core  surrounded by  a
relatively thick CO  buffer whose absolute mass decreases  as the mass
of  the degenerate  ONe core  increases.   Typical masses for this  CO
buffer range from  0.04 to less than $0.01  \, M_{\sun}$, depending on
the mass of the ONe core and on the choice of the prescription adopted
for overshooting.  The  degeneracy in this CO buffer  is also high. On
top of the  CO buffer there is a transition region  in which helium is
most abundant and the degeneracy  rapidly decreases and, on top of it,
a H--rich convective envelope is found.

The  final  outcome of  these  evolutionary  sequences  has been  also
analyzed in detail. Of course, this depends on both the rate of growth
of  the  degenerate  cores   and  on  the  adopted  prescriptions  for
mass--loss.  Moreover, since the  mass--loss rates are most probably a
function  of the  metallicity of  the convective  envelope,  the final
outcome of these stars depends sensitively on this value and, thus, on
the  possible  previous   dredge--up  episodes.   In  particular,  for
non--primordial  stars such  question  would not  represent a  problem
since stellar winds during  the TP--(S)AGB stage are naturally pointed
out  as the mechanism  by which  these stars  are deprived  from their
hydrogen--rich envelopes and, therefore, the formation of white dwarfs
is immediately  inferred.  Such stellar winds can  be easily justified
by the obvious presence of  metals in the envelopes of non--primordial
stars.  However, only the  lightest elements are originally present in
$Z=0$ stars.   Consequently, for the case of  the massive intermediate
mass stars that we have analyzed in this paper the situation is not so
clear.   Their  evolution   through  core  hydrogen--,  helium--  and,
eventually,  carbon--burning,   and  the  dredge--up   processes  that
accompany the TP--(S)AGB phase yield  a modest enrichment in metals of
the surface.  Whether or not  this enrichment is enough to enhance the
mass--loss rates is unclear.  For  instance, we have found that in the
case of  the $7.0\, M_{\sun}$ star  computed disregarding overshooting
the surface abundance  by mass of metals is of  the order of $10^{-5}$
after the $15^{\rm th}$ pulse of the TP--AGB phase whereas in the case
of  the $9.0  \, M_{\sun}$  star the  surface abundance  of  metals is
already of  the order of $10^{-4}$  by the end  of the carbon--burning
phase (Gil--Pons  et al.  2005).  The evolution  during the TP--(S)AGB
phase  of  the  models   computed  with  overshooting  leads  to  more
metal--enriched envelopes because there exists a deeper penetration of
the convective envelope to the regions of the star where core elements
have been synthesized and, therefore,  mass loss is more favored.  In
spite of these metal enrichments  we have shown that if the Schr\"oder
\&  Cuntz  (2005)  prescription  (corrected  for  the  effect  of  the
metallicity  of the  envelope)  is adopted  for  the mass--loss  rates
during  the TP-(S)AGB  phase, all  the models  studied here  seem more
likely to end up their lives  as supernovae.  The reason is that their
cores are able  to grow up to $M_{\rm Ch}$ in  time scales between two
and three orders  of magnitudes shorter than those  required for these
stars to get  rid of their envelopes, in  agreement with the arguments
of Zijlstra (2004).

Taking the prescription of  Bl\"ocker (1995) into account still allows
the  formation of SNI1/2  as a  consequence of  the evolution  of ZAMS
objects   between   5   and   $7   \,   M_{\sun}$   computed   without
overshooting. This  prescription is  very often considered  to produce
very high  mass--loss rates and,  therefore, the fact that  even under
these conditions  the possibility for the formation  of SNI1/2 remains
open is an encouraging result  which deserves further scrutiny. In any
case,  it must be  kept in  mind that  the lack  of a  first principle
theory of mass--loss results  in a considerable complication for these
studies, even in the case of solar metallicity stars.  The conclusions
on the subject must rely  on detailed observations of open clusters or
detached binary systems  in which one of the members  is a white dwarf
--- see  Weidemann   (2000)  for  a   thorough  review  ---   and  any
extrapolation   to   primordial  objects   might   be  still   somehow
speculative.

\begin{acknowledgements}
Part   of    this   work   was    supported   by   the    MCYT   grant
AYA05--08013--C03--01, by  the European Union FEDER funds,  and by the
AGAUR.
\end{acknowledgements}


\end{document}